\documentclass[aps,floats,11pt, nofootinbib,prd,showkeys,superscriptaddress]{revtex4-2}

\usepackage{graphicx}
\usepackage{dcolumn}
\usepackage{bm}

\usepackage{epsfig}
\usepackage{epstopdf}
\usepackage[latin1]{inputenc}
\usepackage{amsmath}
\usepackage{amsfonts}
\usepackage{hyperref}
\usepackage{xcolor}

\def\beq{\begin{equation}}
\def\eeq{\end{equation}}

\def\beqstr{\begin{equation*}}
\def\eeqstr{\end{equation*}}

\def\ber{\begin{eqnarray}}
\def\eer{\end{eqnarray}}

\def\berstr{\begin{eqnarray*}}
\def\eerstr{\end{eqnarray*}}

\def\benu{\begin{enumerate}}
\def\eenu{\end{enumerate}}

\def\l{\left}
\def\r{\right}

\newcommand{\al}{\alpha}
\newcommand{\bet}{\beta}
\newcommand{\el}{\mathcal{L}}

\def\be{\begin{equation}}   
\def\ee{\end{equation}}  
 \def\bea{\begin{eqnarray}}    \def\eea{\end{eqnarray}}      
                    \def\be{\beta}   
\def\r{\right}            \def\l{\left}

\begin{document}

\title{Unified Dark Matter and Dark Energy in a model of Non-Canonical Scalar-Tensor Theory}%

\author{Nihal Jalal Pullisseri}\email{nihaljalal.mathphys@gmail.com}
\affiliation{Universit\"at Leipzig, 04109 Leipzig, Germany}
\author{Sanil Unnikrishnan}\email{u.sanil@ststephens.edu}
\affiliation{Department of Physics, St.\ Stephen's College, University of Delhi, Delhi 110007, India}

\date{\today}
\begin{abstract}
We consider a model of non-canonical scalar-tensor theory in which the kinetic term in the Brans-Dicke action is replaced by a non-canonical scalar field Lagrangian $\mathcal{L}(X, \phi)= \lambda X^\alpha \phi^\beta - V(\phi)$ where $X = (1/2) \partial_{\mu} \phi \partial^{\mu} \phi$ and $\alpha$, $\beta$ and $\lambda$ are parameters of the model.
This can be considered as a simple non-canonical generalization of the Brans-Dicke theory with a potential term which corresponds to a special case of this model with the values of the parameter $\alpha = 1$, $\beta = -1$ and $\lambda = 2w_{_{BD}}$ where $w_{_{BD}}$ is the Brans-Dicke parameter.
Considering a spatially flat Friedmann-Robertson-Walker Universe with scale factor $a(t)$, it is shown that, in the matter free Universe, the kinetic term $\lambda X^\alpha \phi^\beta$ can lead to a power law solution $a(t)\propto t^{n}$ but the maximum possible value of $n$ turns out to be $(1+\sqrt{3})/4 \approx 0.683$.
When $\alpha \geq 18$, this model can lead to a solution $a(t)\propto t^{2/3}$, thereby mimicking the evolution of scale factor in a cold dark matter dominated epoch with Einstein's General Relativity (GR).
With the addition of a linear potential term $V(\phi) = V_{0}\phi$, it is shown that this model mimics the standard $\Lambda$CDM model type evolution of the Universe.
The larger the value of $\alpha$, the closer the evolution of $a(t)$ in this model to that in the $\Lambda$CDM model based on Einstein's GR.
The purpose of this paper is to demonstrate that this model with a linear potential can mimic the GR based $\Lambda$CDM model. 
However, with an appropriate choice of the potential $V(\phi)$, this model can provide a unified description of both dark matter and dynamical dark energy, as if it were based on Einstein's GR.
\end{abstract}

\maketitle
\flushbottom

\section{Introduction}\label{Sec: Introdiction}

Cosmological observations, such as the light curves of Type Ia supernovae and cosmic microwave background radiation, Baryon Acoustic Oscillation (BAO) data, and the large scale structure formation data strongly indicate that the Universe is currently undergoing accelerated expansion~\cite{SupernovaSearchTeam:1998fmf,SupernovaCosmologyProject:1998vns,DESI:2024mwx,DESI:2024mwx,DiValentino:2020evt,Planck:2013pxb,Planck:2018vyg,Planck:2018nkj,SDSS:2005xqv}. 
Within the framework of Einstein's General Theory of Relativity, this observed acceleration implies that approximately $70\%$ of the total content of the Universe at the present epoch is dark energy, which is defined as a form of matter with negative pressure and the ratio of pressure to its energy density is less than $-1/3$~\cite{Copeland:2006wr,Sahni:1999gb,Padmanabhan:2002ji,Frieman:2008sn,Peebles:2002gy,Li:2011sd,Bamba:2012cp}.
Furthermore, cosmological observations also suggest that about $26\%$ of the Universe's total content is dark matter, which behaves as pressureless matter~\cite{Cirelli:2024ssz,Sahni:2004ai,Bertone:2016nfn}. 
The remaining $4\%$ consists of ordinary matter, the types of matter detectable in laboratories, primarily composed of Baryons.

Although numerous models of dark matter and dark energy have been proposed, the fundamental nature of both remains unknown. 
Like ordinary matter, dark matter behaves gravitationally as an attractive form of matter, playing a pivotal role in the formation of large-scale structures in the Universe. 
In contrast, dark energy acts as a repulsive form of matter, driving the observed accelerated expansion of the Universe at late times. Despite their differing gravitational behaviors, the energy densities of both dark matter and dark energy at the present epoch are of similar order of magnitude~\cite{Zlatev:1998tr}. 
This suggests the possibility that dark matter and dark energy could arise from a single matter content, which behaves differently at different epochs and cosmological scales. 
Models of a unified dark sector have also been explored in the literature~\cite{Scherrer:2004au,Bertacca:2007ux,Bertacca:2010ct,Bose:2008ew,Sahni:2015hbf,Mishra:2018tki,Benisty:2018qed,Benisty:2018oyy,Benisty:2017eqh}.

Considering dark matter and dark energy as separate entities, one of the simplest models to explain the late-time accelerated expansion of the Universe is to add a cosmological term $\Lambda$, in Einstein's equation~\cite{Weinberg:1988cp,Carroll:2000fy}. 
Together with pressureless dark matter, also known as cold dark matter, this model with the cosmological constant is known as the $\Lambda$CDM model.
Although recent cosmological observations, such as the Hubble tension and Dark Energy Survey Instrument (DESI) Baryon Acoustic Oscillation (BAO) observational data~\cite{Dainotti:2023yrk,Perivolaropoulos:2021jda,Scolnic:2024hbh,Verde:2023lmm,Pourojaghi:2024bxa,DiValentino:2021izs}, may indicate a deviation from the cosmological expansion predicted by the $\Lambda$CDM model, this simple concordance model has still not been ruled out~\cite{Blanchard:2022xkk,Colgain:2024xqj,Cortes:2024lgw}. 
For the cosmological constant, the equation of state parameter, which is defined as the ratio of pressure to energy density, turns out to be  $-1$.

In addition to the non-zero cosmological constant as an explanation for the late-time accelerated expansion of the Universe, models of dark energy based on scalar fields have also been proposed in the literature. These include quintessence, Tachyon, phantom, and k-essence models of dark energy~\cite{Tsujikawa:2013fta,Nojiri:2005pu,Padmanabhan:2002cp,Caldwell:1999ew,Capozziello:2005tf,Armendariz-Picon:2000ulo,Chiba:1999ka}. 
While quintessence corresponds to scalar field dark energy whose Lagrangian has the standard structure of a kinetic term minus a potential term, k-essence corresponds to non-canonical scalar field models of dark energy, where the Lagrangian is a general function of the scalar field and the kinetic term~\cite{Garriga:1999vw}. 
The equation of state parameter in all these types of scalar field models of dark energy deviates from $-1$ and can also evolve with time.

The existence of dark energy driving the late-time accelerated expansion of the Universe is based on the assumption that gravity on cosmological scales operates according to Einstein's GR. 
However, it is also possible that this assumption may be incorrect, and that the gravitational laws on cosmological scales deviate from those described by GR. 
Various models of modified gravity that account for the late-time accelerated expansion of the Universe have been explored in the literature~\cite{Carroll:2003wy,Nojiri:2006ri,Clifton:2011jh,Koyama:2015vza,Nojiri:2010wj,Nojiri:2017ncd}.
These include Brans-Dicke theory, scalar-tensor theory, $f(R)$ gravity models, and others~\cite{Caldwell:2009ix,Sotiriou:2008rp,DeFelice:2010aj,Faraoni:2010pgm,Capozziello:2011et,Elizalde:2008yf,Hussain:2024yee,Hussain:2025vbo,Arora:2025ecj}.

The Brans-Dicke theory is a class of scalar-tensor theories in which Newton's gravitational constant,  $G_{_{N}}$, is not a constant but a dynamic quantity whose evolution is influenced by the matter content of the Universe~\cite{Brans:1961sx}. 
In this framework, $G_{_{N}} \propto 1/\phi$, where $\phi$ represents the Brans-Dicke field, and its evolution depends on both the Brans-Dicke parameter, $w_{_{BD}}$, and the energy-momentum tensor of the matter content.
The Brans-Dicke theory with a cosmological constant term has been studied to alleviate the Hubble Tension~\cite{SolaPeracaula:2020vpg}. 
A general canonical scalar-tensor theory can be seen as an extension of the Brans-Dicke theory, where the Brans-Dicke parameter is a function of the field $\phi$, i.e., $w_{_{BD}}(\phi)$, and a potential term, $V(\phi)$, is added to the Brans-Dicke action~\cite{Fujii:2003pa,Faraoni:2004pi,Elizalde:2004mq}.
Modified gravity models, such as $f(R)$ gravity, can also be described as a canonical scalar-tensor theory with $w_{_{BD}}(\phi) = 0$, but with a non-zero potential term, $V(\phi)$~\cite{Chiba:2003ir}.
The scalar-tensor theories are referred to as such because they extend beyond the traditional framework of Einstein's gravity, which solely uses the metric tensor to describe gravitational interactions. 
In contrast, Scalar-Tensor theories introduce additional scalar fields along with the metric tensor to describe the gravitational field.

A non-canonical scalar-tensor theory generalizes the canonical scalar-tensor theory by replacing its standard kinetic term with a more general Lagrangian $\mathcal{L}(X, \phi)$, where $X = (1/2) \partial_{\mu} \phi \partial^{\mu} \phi$ represents the kinetic term. 
This is equivalent to modifying the kinetic term in Brans-Dicke theory by replacing it with $\mathcal{L}(X, \phi)$. 
Models with a tachyonic form of $\mathcal{L}(X, \phi)$ have been shown to drive the accelerated expansion of the Universe~\cite{Unnikrishnan:2005wu}, and other models with different forms of $\mathcal{L}(X, \phi)$ have also been explored~\cite{Ferrari:2025egk,Jana:2017hqw,Quiros:2019ktw,Kim:2004is}.

In this paper, we introduce a model of non-canonical scalar-tensor theory in which the usual kinetic term in the Brans-Dicke action is replaced by a term proportional to $X^\alpha \phi^\beta$, where $\alpha$ and $\beta$ are constants. 
This can be considered as a generalization of the Brans-Dicke theory, which corresponds to a special case of this non-canonical scalar-tensor theory with parameters $\alpha$ and $\beta$ set to $1$ and $-1$, respectively.
Considering a spatially flat Friedman-Robertson-Walker Universe, we show that this non-canonical scalar-tensor model admits a power law solution for the scale factor $a(t) \propto t^n$, along with a similar solution for the scalar field, i.e., $\phi(t) \propto t^m$, in a matter free Universe. 
However, in this model, it turns out that the maximum value that $n$ in the solution $a(t) \propto t^n$ can take is $0.68$. 
Therefore, the matter free case in this model cannot lead to accelerated expansion of the Universe, which requires $n > 1$. 
However, $n = 2/3$ is a possible solution in this model, which corresponds to the solution for the scale factor $a(t)$ in Einstein's GR when the Universe is dominated by pressureless matter.
Further, by adding a linear potential term $V(\phi)$ to the action, we show that this model mimics the evolution of $a(t)$ as if it were in the $\Lambda$CDM model in Einstein's GR. The greater the value of the parameter $\alpha$, the closer the solution $a(t)$ in this model approaches the corresponding $a(t)$ in the GR $\Lambda$CDM model.
Notably, in our model, the scalar field in the non-canonical scalar-tensor theory behaves like cold dark matter at early times and like $\Lambda$ (dark energy) at late times, providing a unified description of the dark sector.

This paper is organized as follows. 
In Sec.~(\ref{Sec: Canonical STT}) we review the canonical scalar-tensor theory, which is equivalent to Brans-Dicke theory with a potential $V(\phi)$ and with a scalar field dependent Brans-Dicke parameter $w_{_{BD}}(\phi)$.
We then generalize this to a non-canonical scalar-tensor theory in  Sec.~(\ref{Sec: General NC STT}).
A specific model of non-canonical scalar-tensor theory and its cosmological evolution equations are introduced in  Sec.~(\ref{Sec: Our NC STT Model}).
Further, in  Sec.~(\ref{Sec: power law evolution}), it is shown that the kinetic term in the model introduced in the Sec.~(\ref{Sec: Our NC STT Model}) leads to a power law solution for the scale factor. 
In addition to the kinetic term, a linear potential term is considered in the model and it is shown in Sec.~(\ref{Sec: LCDM evolution}) that this model of non-canonical scalar-tensor theory mimic the evolution of the scale factor as if it were a $\Lambda$CDM model based on Einstein's GR.
We summarize the main results of this paper in Sec.~(\ref{Sec: conclusions}).

\section{Canonical Scalar-tensor Theory}\label{Sec: Canonical STT}
The action in a general canonical scalar-tensor theory is given by
\begin{equation}
\mathcal{S} = \frac{1}{16 \pi} \int \l[-f(\psi)R + \frac{1}{2} \partial_\mu \psi \partial^{\mu} \psi - U(\psi)\r]\sqrt{-g}\, d^4 x + \int \el_m \sqrt{-g}\,d^4 x, \label{Eqn: Canonical STA}
\end{equation}
where $U(\psi)$ is the scalar field potential and $(1/2)\partial_a \psi \partial^{a} \psi$ is the kinetic term and $\el_m$ is the Lagrangian of the matter.
The above action replaces the gravity part of the standard Einstein-Hilbert action namely $(1/16 \pi G)\l[-R\r]$ with the term $(1/16 \pi)\l[-f(\psi)R + (1/2)\partial_\mu \psi \partial^{\mu} \psi - U(\psi)\r]$. Thus, in the scalar-tensor theory of gravity, not only the metric tensor but also the value of the scalar field determines the gravitational interactions. Note that, compared with the standard Einstein-Hilbert action, the above action for the scalar-tensor theory also implies that the Newtonian gravitational constant $G = 1/f(\psi)$ is no longer constant, but instead depends on the scalar field, which may evolve with time.

Without loss of generality, we can introduce a field $\phi$ defined as
\begin{equation*}
\phi = f(\psi),
\end{equation*}
where $f $ is a smooth and invertible function. 
Using this, the above action~(\ref{Eqn: Canonical STA}) becomes
\begin{equation}
\mathcal{S} = \frac{1}{16\pi} \int \l[- \phi R + \omega_{_{BD}}(\phi)\l(\frac{\partial_{\mu} \phi \partial^{\mu} \phi}{\phi}\r)  - V(\phi)\r] \sqrt{-g}\,d^4 x + \int \el_m \sqrt{-g}\,d^4 x, \label{Eqn: generalized BDT}
\end{equation}
where 
$$ \omega_{_{BD}}(\phi) = \frac{f(\psi)}{2f'(\psi)}~~~\mathrm{and}~~~~V(\phi) = U(\psi)$$
where $\psi$ can be expressed as a function of $\phi$ by inverting the function $\phi = f(\psi)$. 
A prototype of the scalar-tensor theory called the Brans-Dicke theory corresponds to the case when in the action~(\ref{Eqn: generalized BDT}), $\omega_{_{BD}}(\phi) = \omega_{_{BD}} = \mathrm{constant}$ and $V(\phi) = 0$. 
With this, the action for the Brans-Dicke theory is given by~\cite{Brans:1961sx}
\begin{equation}
\mathcal{S}_{_{\mathrm{BD}}} = \frac{1}{16\pi} \int \l[- \phi R + \omega_{_{BD}}\l(\frac{\partial_{\mu} \phi \partial^{\mu} \phi}{\phi}\r)\r] \sqrt{-g}\,d^4 x + \int \el_m \sqrt{-g}\,d^4 x. \label{Eqn: Action BDT}
\end{equation}
In Brans-Dicke theory, the Newtonian gravitational constant $G = 1/\phi$ is not a constant. 
Note that dividing the kinetic term in the Brans-Dicke action by $\phi$ makes the Brans-Dicke parameter  $\omega_{_{BD}}$ a dimensionless constant. 

Comparing the standard Brans-Dicke theory represented by the action~(\ref{Eqn: Action BDT}) with the action for the general canonical scalar-tensor theory given in Eq.~(\ref{Eqn: generalized BDT}), it is evident that action~(\ref{Eqn: generalized BDT}) can be considered as the generalized Brans-Dicke theory where its parameter $\omega_{_{BD}}(\phi)$ is no longer a constant and the evolution of $\phi$ is also governed by a non-zero potential term $ V(\phi)$.
Note that since action~(\ref{Eqn: generalized BDT}) is an equivalent representation of action~(\ref{Eqn: Canonical STA}), the canonical scalar-tensor theory represented by action~(\ref{Eqn: generalized BDT}) is indeed a canonical generalization of the Brans-Dicke theory. 
Even an $f(R)$ model of modified gravity can be equivalently represented by such a generalized Brans-Dicke theory with its parameter $\omega_{_{BD}}(\phi) = 0$ but with a non-zero potential term $ V(\phi)$ in the action~(\ref{Eqn: generalized BDT})~\cite{Chiba:2003ir,Olmo:2006eh}.

\section{General Non-Canonical scalar-tensor Theory}\label{Sec: General NC STT}
The action for a general non-canonical scalar-tensor theory with a non-canonical scalar field can be expressed as  
\begin{equation}
\mathcal{S} = \frac{1}{16 \pi} \int \l[-f(\psi)R + p(Y, \psi)\r]\sqrt{-g}\, d^4 x + \int \el_m \sqrt{-g}\,d^4 x, 
\label{Eqn: Action NC STT-1}
\end{equation}
where $ p(Y, \psi)$ is the Lagrangian of the non-canonical scalar field $\psi$ where $Y = (1/2)\partial_{\mu} \psi \partial^{\mu} \psi$ is the kinetic term.
Within the framework of Einstein's GR, a non-canonical scalar field can drive inflation~\cite{Unnikrishnan:2012zu,Armendariz-Picon:1999hyi,Chen:2006nt} and it can also drive the late-time accelerated expansion of the Universe as a model of dark energy~\cite{Unnikrishnan:2008ki, Armendariz-Picon:2000nqq,Malquarti:2003nn,Csillag:2025gnz}.
In this paper, we consider a non-canonical scalar field within the framework of scalar-tensor theories in which the non-canonical scalar field is non-minimally coupled to the Ricci scalar in the Einstein-Hilbert action and to investigate the possibility of such a scalar field behaving as both dark matter and dark energy.
A non-canonical scalar field which is non-minimally coupled to gravity can also drive inflation~\cite{Oikonomou:2021edm}.

Note that the action~(\ref{Eqn: Action NC STT-1}) is the non-canonical generalization of the action~(\ref{Eqn: Canonical STA}) with the standard canonical scalar field Lagrangian in the action~(\ref{Eqn: Canonical STA})  \emph{viz}..\ $\mathcal{L} = Y - V(\psi)$ being replaced by a non-canonical Lagrangian $ p(Y, \psi)$.
Similar to the analysis done in the preceding section and without loss of generality, we can introduce a new scalar field defined as $\phi = f(\psi)$.
With this, the action~~(\ref{Eqn: Action NC STT-1}) becomes
\begin{equation}
\mathcal{S} = \frac{1}{16\pi} \int \left[ -\phi R + \el(X, \phi) \right] \sqrt{-g}\,d^4 x + \int \el_m \sqrt{-g}\, d^4 x,
\label{Eqn: Action NC STT-2}
\end{equation}
where $\el(X, \phi)$ is the Lagrangian of the scalar field $\phi$, which is a general function of $\phi$ and the kinetic term $X$ is defined as 
$$X = \frac{1}{2}\partial_{\mu} \phi \partial^{\mu} \phi .$$
In the above action~(\ref{Eqn: Action NC STT-2}), the Lagrangian $\el(X, \phi)$ is such that $\el(X, \phi)\,=\, p\l(Y(X, \phi)\, ,\, \psi(\phi)\r)$ which is obtained by inverting the function $\phi = f(\psi)$ and substituting this inverted function in the action~(\ref{Eqn: Action NC STT-1}).

Note that the standard Brans-Dicke theory described by the action~(\ref{Eqn: Action BDT}) corresponds to a special case of action~(\ref{Eqn: Action NC STT-2}) with $\el(X, \phi) = 2\,\omega_{_{BD}}X/\phi$. 
Further, the generalized Brans-Dicke theory described by the action~(\ref{Eqn: generalized BDT}) corresponds to the case when, in the action~(\ref{Eqn: Action NC STT-2}), $\el(X, \phi) = 2\,\omega_{_{BD}}(\phi)X/\phi\,-\, V(\phi)$.
Therefore, the action~(\ref{Eqn: Action NC STT-2}) or equivalently the action~~(\ref{Eqn: Action NC STT-1}) can be viewed as the non-canonical generalization of the standard Brans-Dicke theory.

In this paper, we focus on a particular form of the Lagrangian $\el(X, \phi)$ in the action~(\ref{Eqn: Action NC STT-2}) which can lead to a $\Lambda$CDM like evolution of the scale factor in the absence of matter. In this scenario, the same scalar field can mimic the behavior of both cold dark matter and dark energy. Before proceeding to the specific model, we first describe in this section all the relevant equations for the general non-canonical scalar-tensor theory defined by the action.

On varying the action~(\ref{Eqn: Action NC STT-2}) with respect to the metric tensor $g_{\mu\nu}$ leads to the following equations
\begin{equation} 
 G_{\mu\nu} = \l(\frac{8 \pi}{\phi}\r) T_{\mu\nu} \,+\, \frac{1}{\phi}  \l(\phi_{,\mu\, ; \nu}\,-\,g_{\mu\nu} \Box \phi\r) \,+\,\frac{1}{2\phi}\left( \el_X \partial_{\mu} \phi \partial_{\nu} \phi - \el g_{\mu\nu}\right), 
 \label{Eqn: Einsten Eqn general NC STT}
 \end{equation}
where $G_{\mu\nu}$ is the Einstein-tensor, $T_{\mu\nu}$ is the energy-momentum tensor of the matter content described by the Lagrangian $\el_m$ in the action~(\ref{Eqn: Action NC STT-2}) which is defined as
\begin{equation}
T^{\mu\nu}\,=\, -\l(\frac{2}{\sqrt{-g}}\r)\l(\frac{\delta}{\delta g_{\mu\nu}}\l(\el_m\sqrt{-g}\r)\r).\label{Eqn: EM Tensor for matter}
\end{equation}
Note that in Eq.~(\ref{Eqn: Einsten Eqn general NC STT}), $\el_X$ is the partial derivative of Lagrangian $\el(X, \phi)$ with respect to the kinetic term $X$.
On varying the action~(\ref{Eqn: Action NC STT-2}) with respect to the scalar field $\phi$ leads to the following equation of motion for $\phi$
\begin{equation} 
\el_X \Box \phi \,=\, \el_{\phi} - 2X \el_{X \phi} - \el_{XX} \partial_{\mu}X\partial^{\mu}\phi - R,
\label{Eqn: Field eqn phi GNC STT}
\end{equation}
where $R$ is the Ricci scalar. 
By taking the trace of equation~(\ref{Eqn: Einsten Eqn general NC STT}), we can get an expression for the Ricci scalar $R$ which on substituting in Eq.~(\ref{Eqn: Field eqn phi GNC STT}) gives
\begin{equation} 
 \Box \phi = \l(\frac{8\pi T}{\phi \el_{X}+3}\r) \,+\, \l(\frac{\phi \el_{\phi} + X \el_{X} - 2X\phi \el_{X \phi} -2\el-\phi\el_{XX} \partial_{\mu}X\partial^{\mu}\phi }{\phi \el_{X}+3}\r),
 \label{Eqn: box phi eqn phi GNC STT}
\end{equation}
where $T$ is the trace of the energy-momentum tensor of the matter content described by the Lagrangian $\el_m$ in the action~(\ref{Eqn: Action NC STT-2}).

We consider a spatially flat Friedmann-Robertson-Walker (FRW) Universe with the line element given by
\begin{equation}
ds^{2} = dt^{2} - a^{2}(t)[dx^{2}+dy^{2}+dz^{2}],\label{Eqn: FRW line element}
\end{equation}
where $a(t)$ is the scale factor. 
The evolution of the scale factor is governed by Einstein's equations~(\ref{Eqn: Einsten Eqn general NC STT}), which leads to the following two Friedmann's equations
\begin{eqnarray}
\frac{\dot{a}^2}{a^2} &=&   \l(\frac{8\pi }{3 \phi}\r)\rho\, - \,H\l(\frac{\dot \phi}{\phi}\r) \,+\, \frac{1}{6\phi}\left[ 2X \el_X - \el \right],  \label{Eqn: Friedman Eqn-1 GNC-STT}\\
\frac{\ddot{a}}{a} &=&    -\l(\frac{4 \pi}{3 \phi}\r)\l(\rho+3p\r)\, - \,\frac{\ddot \phi}{2 \phi}\, - \,\frac{H \dot\phi}{2\phi}\, - \,\frac{1}{6\phi}\left[ X \el_X + \el \right],
\label{Eqn: Friedman Eqn-2 GNC-STT}
\end{eqnarray}
where $\rho$ and $p$ are the energy density and pressure of the matter content, respectively, as determined by the energy-momentum tensor $T^{\mu\nu}$ given in Eq.~(\ref{Eqn: EM Tensor for matter}). 
Note that in the FRW line element, the energy-momentum tensor can be expressed as $T^{\mu}_{\;\;\nu} = \mathrm{diag}\l[\rho,\, -p,\,-p,\,-p\r]$.
The conservation of the energy-momentum tensor $T^{\mu}_{\;\;\nu\, ;\,\mu}\,=\, 0$ implies that $\dot{\rho}=-3H(\rho + p)$.

In the FRW space-time with line-element~(\ref{Eqn: FRW line element}), the scalar field $\phi$ is a function only of time, i.e $\phi = \phi(t)$. 
Consequently, the field equation for $\phi$ as described by Eq.~(\ref{Eqn: box phi eqn phi GNC STT}) becomes
\begin{equation} 
\ddot \phi = \frac{8\pi(\rho-3p)-3H\dot\phi(\phi \el_X +3) +( \phi \el_{\phi}+X\el_X -2X\phi \el_{X \phi}-2\el)}{2X\phi \el_{X X} + \phi\el_X + 3}.\label{Eqn: phi dot dot GNC STT}
\end{equation}
The equation above for $\phi$ and the Friedmann equation~(\ref{Eqn: Friedman Eqn-1 GNC-STT}) along with the conservation equations $\dot{\rho}=-3H(\rho + p)$ for each matter content, form a closed set of equations for determining the evolution of scale factor $a(t)$.

Note that the standard Brans-Dicke theory corresponds to the case when $\el(X, \phi) = 2\omega_{_{BD}}X/\phi$ in the action~(\ref{Eqn: Action NC STT-2}).
Substituting this in Eqs.~(\ref{Eqn: Friedman Eqn-1 GNC-STT}), (\ref{Eqn: Friedman Eqn-2 GNC-STT}), and (\ref{Eqn: phi dot dot GNC STT}) give the following Friedmann equations and the field equation for $\phi$ in the Brans-Dicke theory
\begin{eqnarray}
\frac{\dot{a}^2}{a^2} &=&   \l(\frac{8\pi }{3 \phi}\r)\rho\,+\, \frac{\omega_{_{BD}}}{6}\l(\frac{\dot{\phi}}{\phi}\r)^{2}\, - \,H\l(\frac{\dot \phi}{\phi}\r), \label{Eqn: Friedman Eqn-1 BDT}\\
\frac{\ddot{a}}{a} &=&    -\l(\frac{4 \pi}{3 \phi}\r)\l(\rho+3p\r) \,-\, \frac{\omega_{_{BD}}}{3}\l(\frac{\dot{\phi}}{\phi}\r)^{2}  \, - \,\frac{\ddot \phi}{2 \phi}\, - \,\frac{H \dot\phi}{2\phi},
\label{Eqn: Friedman Eqn-2 BDT}\\
\ddot \phi &=& \l(\frac{8\pi}{2\omega_{_{BD}} \,+\, 3}\r)T - 3 H \dot{\phi},\label{Eqn: phi dot dot BDT}
\end{eqnarray}
where $T = \rho-3p$ is the trace of the energy momentum tensor of the matter content.

\section{A model of non-canonical scalar-tensor theory}\label{Sec: Our NC STT Model}
In this paper, we introduce a model of non-canonical scalar-tensor theory with the action given by
\begin{equation}
\mathcal{S} = \frac{1}{16\pi} \int \left[ -\phi R \,+\, \lambda X^{\alpha}\phi^{\beta}\,-\,V(\phi) \right] \sqrt{-g}\,d^4 x + \int \el_m \sqrt{-g}\, d^4 x,
\label{Eqn: Action our new NC STT}
\end{equation}
where $\alpha$, $\beta$ and $\lambda$ are constants or parameters of this model. 
Note that the standard Brans-Dicke theory described by the action~(\ref{Eqn: Action BDT}) corresponds to a special case of the above action~(\ref{Eqn: Action our new NC STT}) with $\alpha = 1$, $\beta = -1$, $\lambda = 2\omega_{_{BD}}$ and $V(\phi) = 0$.
Therefore, the action~(\ref{Eqn: Action our new NC STT}) can be considered as a simple non-canonical generalization of the general Brans-Dicke theory described by the action~(\ref{Eqn: generalized BDT}), which is also equivalent to the canonical scalar-tensor theory~(\ref{Eqn: Canonical STA}). Here, we take $\al$ to be a positive integer and $\lambda$ to be a positive real number.

The action~(\ref{Eqn: Action our new NC STT}) corresponds to choosing the  scalar field Lagrangian $\el(X, \phi)$ in the general non-canonical scalar-tensor theory described by the action~(\ref{Eqn: Action NC STT-2}) as
\begin{equation}
\el(X, \phi)\,=\,\lambda X^{\alpha}\phi^{\beta}\,-\,V(\phi). \label{Eqn: New NC STT Lagrangian}
\end{equation}
With this Lagrangian $\el(X, \phi)$, the Friedmann equations~(\ref{Eqn: Friedman Eqn-1 GNC-STT}) and (\ref{Eqn: Friedman Eqn-2 GNC-STT}) and the field equation for $\phi$  \emph{viz}. Eq.~(\ref{Eqn: phi dot dot GNC STT}) becomes:
\begin{eqnarray}
\frac{\dot{a}^2}{a^2} &=&   \l(\frac{8\pi }{3 \phi}\r)\rho\, - \,H\l(\frac{\dot \phi}{\phi}\r) \,+\, \l(\frac{2\alpha - 1}{6}\r)\lambda X^{\alpha}\phi^{\beta-1}\, +\, \frac{V(\phi)}{6\phi},     \label{Eqn: Friedman Eqn-1 Our GNC-STT}\\
\frac{\ddot{a}}{a} &=&    -\l(\frac{4 \pi}{3 \phi}\r)\l(\rho+3p\r)\, - \,\frac{\ddot \phi}{2 \phi}\, - \,\frac{H \dot\phi}{2\phi}\, - \,\l(\frac{\alpha + 1}{6}\r)\lambda X^{\alpha}\phi^{\beta-1}\, +\, \frac{V(\phi)}{6\phi},     
\label{Eqn: Friedman Eqn-2 Our GNC-STT}\\
\ddot \phi &=& \frac{8\pi(\rho-3p)  -3H\dot\phi(\alpha\lambda X^{\alpha-1}\phi^{\beta+1} + 3) + (\alpha +\beta - 2\alpha\beta -2)\lambda X^{\alpha}\phi^{\beta}}{\alpha (2\alpha -1)\lambda X^{\alpha-1}\phi^{\beta+1} + 3}\nonumber\\ 
     && \; +\; \frac{2V(\phi) - \phi V'(\phi)}
{\alpha (2\alpha -1)\lambda X^{\alpha-1}\phi^{\beta+1} + 3}.\label{Eqn: phi dot dot our GNC STT}
\end{eqnarray}

In this paper, we are interested in investigating the possibility of a $\Lambda$CDM-like evolution of the scale factor $a(t)$ in this model when the contribution of matter is insignificant i.e. when $\el_m = 0$ in the action~(\ref{Eqn: Action our new NC STT}).
This model will then lead to a uniform description of both dark matter and dark energy by the same non-canonical scalar field of the scalar-tensor theory.

We are interested in the solution for $a(t)$ in this model such that it mimics the same solution as one gets in the $\Lambda$CDM model when the gravity is described by Einstein's GR.
Note that in Einstein's GR, the scale factor satisfies the following Friedmann equations in the $\Lambda$CDM model
\begin{eqnarray}
\frac{\dot{a}^2}{a^2} &=&   \l(\frac{8\pi G }{3}\r)\rho_{_{m0}}a^{-3}\, + \,  \frac{\Lambda}{3},   \label{Eqn: Friedman Eqn-1 Lamda CDM}\\
\frac{\ddot{a}}{a} &=&    -\l(\frac{4 \pi G}{3}\r)\rho_{_{m0}}a^{-3}\, + \,  \frac{\Lambda}{3}, 
\label{Eqn: Friedman Eqn-2 Lambda CDM}
\end{eqnarray}
where $\rho_{_{m0}}$ is the energy density of the cold dark matter at the present epoch, at which $a(t) = 1$ and $\Lambda$ is the cosmological constant.
Comparing Eqs.~(\ref{Eqn: Friedman Eqn-1 Lamda CDM}) and (\ref{Eqn: Friedman Eqn-2 Lambda CDM}) with Eqs.~(\ref{Eqn: Friedman Eqn-1 Our GNC-STT}) and (\ref{Eqn: Friedman Eqn-2 Our GNC-STT}), respectively, it is evident that if we consider a linear potential of the form $V(\phi) = V_0 \phi$ in the action~(\ref{Eqn: Action our new NC STT}), then twice the value of the constant $V_0$ can act as the cosmological constant term in the Friedmann equations~(\ref{Eqn: Friedman Eqn-1 Our GNC-STT}) and (\ref{Eqn: Friedman Eqn-2 Our GNC-STT}). 
However, in the matter free case with $\rho = p = 0$ in Eqs.~(\ref{Eqn: Friedman Eqn-1 Our GNC-STT}) and (\ref{Eqn: Friedman Eqn-2 Our GNC-STT}), to have a $\Lambda$CDM model like solution for $a(t)$, it is required that initially $a(t) \propto t^{2/3}$ to mimic the initial cold dark matter dominated evolution in a similar way as in the case of $\Lambda$CDM model within  Einstein's GR.
This is possible if the kinetic term in the Lagrangian~(\ref{Eqn: New NC STT Lagrangian})  \emph{viz}.\ $\lambda X^{\alpha}\phi^{\beta}$ leads to a solution $a(t) \propto t^{2/3}$ initially when the contribution of the linear potential term with $V(\phi) = V_0 \phi$ is insignificant.
Later, this linear potential term can lead to the accelerated expansion of the Universe, similar to the one driven by the cosmological constant term in Einstein's GR.
In the next section, we will discuss the possibility of having a solution of the type  $a(t) \propto t^{2/3}$ when the kinetic term $\lambda X^{\alpha}\phi^{\beta}$ significantly dominates the potential term in the Lagrangian~(\ref{Eqn: New NC STT Lagrangian}) and with insignificant contribution of matter Lagrangian $\el_m$ in the action~(\ref{Eqn: Action our new NC STT}).

\section{Power law solution when $\el(X, \phi) = \lambda X^{\alpha}\phi^{\beta}$}\label{Sec: power law evolution}
Let us first consider the scenario when the contribution of matter described by the Lagrangian $\el_m$ and the potential term $V(\phi)$ in the action~(\ref{Eqn: Action our new NC STT}) are insignificant. This means that we are considering the case when only the term $\lambda X^{\alpha}\phi^{\beta}$ in the action~(\ref{Eqn: Action our new NC STT}) drives the expansion of the Universe.
With $V(\phi) = 0$ along with $\rho = p = 0$, the Friedmann equations~(\ref{Eqn: Friedman Eqn-1 Our GNC-STT}) and (\ref{Eqn: Friedman Eqn-2 Our GNC-STT}) along with the field equation of $\phi$ given by Eq.~(\ref{Eqn: phi dot dot our GNC STT}) becomes
\begin{eqnarray}
\frac{\dot{a}^2}{a^2} &=&   - H\l(\frac{\dot \phi}{\phi}\r) \,+\, \l(\frac{2\alpha - 1}{6}\r)\lambda X^{\alpha}\phi^{\beta-1},   \label{Eqn: Friedman Eqn-1 power law}\\
\frac{\ddot{a}}{a} &=&    - \frac{\ddot \phi}{2 \phi}\, - \,\frac{H \dot\phi}{2\phi}\, - \,\l(\frac{\alpha + 1}{6}\r)\lambda X^{\alpha}\phi^{\beta-1},
\label{Eqn: Friedman Eqn-2 power law}\\
\ddot \phi &=& \frac{-3H\dot\phi(\alpha\lambda X^{\alpha-1}\phi^{\beta+1} + 3)  + (\alpha + \beta - 2\alpha\beta -2)\lambda X^{\alpha}\phi^{\beta}}
{\alpha (2\alpha -1)\lambda X^{\alpha-1}\phi^{\beta+1} + 3}.
\label{Eqn: phi dot dot power law}
\end{eqnarray}

We are interested in investigating whether the above equations admit a power law solution for both $a(t)$ and $\phi(t)$ and for what values of $\alpha$ and $\beta$ such power law solutions are viable.
We consider the power law solutions of the form
\begin{eqnarray}
a(t) &=&   a_{_i}\l(\frac{t}{t_i}\r)^{n},\label{Eqn: a(t) power law}\\
\phi(t) &=&   \phi_{_i}\l(\frac{t}{t_i}\r)^{m},\label{Eqn: phi(t) power law}
\end{eqnarray}
where $m$ and $n$ are real numbers with $n > 0$ as we consider the expanding Universe solution for $a(t)$.
In Eqs.~(\ref{Eqn: a(t) power law}) and (\ref{Eqn: phi(t) power law}), $a_{_i}$ and $\phi_{_i}$ are the values of the scale factor and scalar field, respectively,  at an initial time $t = t_i$.

From the Friedmann equations~(\ref{Eqn: Friedman Eqn-2 power law}) and (\ref{Eqn: phi dot dot power law}), it follows that 
\begin{equation}
\frac{\lambda X^{\alpha}\phi^{\beta-1}}{2} = -\frac{\ddot \phi}{\phi} - \frac{2H\dot \phi}{\phi} - \l(2\frac{\ddot{a}}{a} + \frac{\dot{a}^2}{a^2}\r). \label{Eqn: K-Term power law}
\end{equation}
Substituting the power law solutions given by Eqs.~(\ref{Eqn: a(t) power law}) and (\ref{Eqn: phi(t) power law}) in the above equation implies that
\begin{equation}
\lambda X^{\alpha}\phi^{\beta-1} \,=\,  \frac{C_1}{t^{2}}, \label{Eqn: K-Term t power -2}
\end{equation}
where the constant $C_1$ is given by
\begin{equation}
C_1 \,=\,  -2\l(3n^{2} + m^{2} +2mn - 2n -m\r).\label{Eqn: C_1}
\end{equation}
With $\phi(t) \propto t^m$, it follows from Eq.~(\ref{Eqn: K-Term t power -2}) that 
\begin{equation}
\beta\,=\,  \frac{ (m-2) \,+\,    2\alpha(1 -m)}{m}.\label{Eqn: beta relation}
\end{equation}
It is evident from the above equation that when $\alpha =1$ one gets $\beta = -1$.
Note that $\alpha =1$ and  $\beta = -1$ in the action~(\ref{Eqn: Action our new NC STT}) with $V(\phi) = 0$ corresponds to the Brans-Dicke theory described by the action~(\ref{Eqn: Action BDT}).
As we are considering the non-canonical scalar-tensor theory, which corresponds to the case when $\alpha \neq 1$, Eq.~(\ref{Eqn: beta relation}) can be expressed as
\begin{equation}
m \,=\,\frac{2(\al-1)}{ 2\al + \bet - 1}.\label{Eqn: m relation on alpha and beta}
\end{equation}
This relation~(\ref{Eqn: m relation on alpha and beta}) gives the value of $m$ in the power law solution $\phi(t) \propto t^m$ for a given value of the parameter $\alpha$ and  $\beta$ in the Lagrangian $\el(X, \phi) = \lambda X^{\alpha}\phi^{\beta}$ with the assumption that $\alpha \neq 1$.

To determine the value of $n$ in the solution $a(t) \propto t^n$ for given values of  $\alpha$ and  $\beta$, we substitute Eqs.~(\ref{Eqn: a(t) power law}) and (\ref{Eqn: phi(t) power law}) in the Friedmann equation~(\ref{Eqn: Friedman Eqn-1 power law}) and this gives the following equation
\begin{equation} 
\alpha \,=\, \frac{1}{2}\l[\frac{m^{2} -m -2n -mn}{3n^{2} -2n + m^{2} -m + 2mn}\r].\label{Eqn: alpha relation on m and n}
\end{equation}
Note that while arriving at Eq.~(\ref{Eqn: K-Term power law}) and the consequent Eq.~(\ref{Eqn: beta relation}), we have combined both the Friedmann equations~(\ref{Eqn: Friedman Eqn-1 power law})
and (\ref{Eqn: Friedman Eqn-2 power law}) whereas Eq.~(\ref{Eqn: alpha relation on m and n}) follows only from the first Friedmann equation~(\ref{Eqn: Friedman Eqn-1 power law}).
It can be verified that on substituting the power law solutions $a(t) \propto t^n$ and $\phi(t) \propto t^m$ in the second Friedmann equation~(\ref{Eqn: Friedman Eqn-2 power law}) one gets the same relations for $\alpha$ as given in Eq.~(\ref{Eqn: alpha relation on m and n}).

On substituting the power law solutions  \emph{viz}..\  $a(t) \propto t^n$ and $\phi(t) \propto t^m$ in the field equation for $\phi$ described by the Eq~(\ref{Eqn: phi dot dot power law}), we get the following relation
\begin{equation} 
    m(m-1) \,=\,\frac{-3mn(2\alpha C_1 + 3 m^{2}) + C_1 m^{2}(\alpha + \beta - 2\alpha\beta -2)}{2\alpha(2\alpha-1)C_1 + 3 m^{2}},\label{Eqn: m(m-1) relation}
\end{equation}
where $C_1$ is given by Eq.~(\ref{Eqn: C_1}).

It is straightforward to verify that Eq.~(\ref{Eqn: m(m-1) relation}) is consistent with Eqs.~(\ref{Eqn: m relation on alpha and beta}) and (\ref{Eqn: alpha relation on m and n}). 
Consequently, Eq.~(\ref{Eqn: m(m-1) relation}) is not an independent relation. Therefore, the two Friedmann equations and the field equation for $\phi$  \emph{viz}..\ Eqs.~(\ref{Eqn: Friedman Eqn-1 power law}) to (\ref{Eqn: phi dot dot power law}) implies that for power law solution of the form $a(t) \propto t^n$ and $\phi(t) \propto t^m$, the two independent equations given by Eqs.~(\ref{Eqn: m relation on alpha and beta}) and (\ref{Eqn: alpha relation on m and n}) must be satisfied using which the values of $m$ and $n$ can be determined for given values of the parameter $\alpha$ and  $\beta$ in the Lagrangian $\el(X, \phi) = \lambda X^{\alpha}\phi^{\beta}$.

To determine the value of $n$ in the power law solution $a(t) \propto t^n$, we use Eq.~(\ref{Eqn: alpha relation on m and n}) together with Eq.~(\ref{Eqn: m relation on alpha and beta}) for $m$.
Consequently, Eq.~(\ref{Eqn: alpha relation on m and n}) can be expressed as a quadratic equation on $n$ which leads to the following roots for $n$
\begin{equation}
n =\frac{2 - \beta + \alpha (2 \beta - 1) \pm \sqrt{S}}{6\alpha(2\alpha + \beta - 1)},\label{Eqn: n relation on alpha and beta}
\end{equation}
where 
\begin{equation}
S \,=\, (\beta - 2)^2 + 24 \alpha^3 (1 + \beta) + \alpha (8 + 22 \beta - 4 \beta^2) + \alpha^2 (-35 - 40 \beta + 4 \beta^2).\label{Eqn: S main eqn}
\end{equation}

For $n$ to be a real number, it is required that $S\geq 0$. 
Note that, for a given value of $\alpha$, Eq.~(\ref{Eqn: S main eqn}) can be expressed as a quadratic equation in $\beta$ in the following way
\begin{equation}
S \,=\, \mathcal{P}_{_2} \beta^{2} + \mathcal{P}_{_1} \beta + \mathcal{P}_{_0},\label{Eqn: S quadratic eqn}
\end{equation}
where
\begin{eqnarray}
\mathcal{P}_{_2}\, &=& \,(2\al -1)^2, \label{Eqn: P1}\\
\mathcal{P}_{_1}\, &=& \, 24 \alpha^{3} - 40 \alpha^{2}  + 22\alpha - 4,\label{Eqn: P2}\\
\mathcal{P}_{_0}\, &=& \, 24 \alpha^{3} - 35 \alpha^{2}  + 8\alpha + 4.\label{Eqn: P3}
\end{eqnarray}
Note that $\mathcal{P}_{_2} > 0$ for $\alpha > 1$.
Therefore, for a given value of $\alpha$, Eq.~(\ref{Eqn: S quadratic eqn}) implies that $S$ is a quadratic function of $\beta$ with its parabola opening upwards.
This means that, for $S \geq 0$, the value of $\beta$ must be such that 
\begin{equation}
\beta \, \leq \, \frac{-\mathcal{P}_{_1} - \sqrt{\mathcal{P}_{_1}^2 - 4 \mathcal{P}_{_2}\mathcal{P}_{_0}}}{2\mathcal{P}_{_2}}~~~~~~~\mathrm{or}~~~~~~~\beta \, \geq \, \frac{-\mathcal{P}_{_1} + \sqrt{\mathcal{P}_{_1}^2 - 4 \mathcal{P}_{_2}\mathcal{P}_{_0}}}{2\mathcal{P}_{_2}}.\label{Eqn: beta range 1}
\end{equation}
Consequently, for $n$ to be a real number, the value of $\beta$ for each value of $\alpha > 1$ must be such that 
\begin{equation}
\beta \, \notin \, \l(B\, ,\,B'\r), \label{Eqn: beta forbidden range 1}
\end{equation}
where
\begin{eqnarray}
B\, &=& \, -3\alpha \,-\, 6\sqrt{\frac{(\alpha - 1)^3 \alpha}{(2\alpha - 1)^2}} \,+\, 2, \label{Eqn: B1}\\
B'\, &=& \,-3\alpha \,+\, 6\sqrt{\frac{(\alpha - 1)^3 \alpha}{(2\alpha - 1)^2}} \,+\, 2.   \label{Eqn: B2}
\end{eqnarray}
This forbidden open interval $\l(B\, ,\,B'\r)$ for the value of the $\beta$ is a necessary condition for having a real value of $n$ in the power law solution $a(t) \propto t^{n}$.
Note that we are considering the non-canonical scalar-tensor theory with $\alpha$ being a positive integer greater than one.
It is important to emphasize that, for $\alpha > 1$, the value of $(1 - 2\alpha)$ always lies within the open interval $\l(B\, ,\,B'\r)$. 
Therefore, for $\alpha > 1$, the allowed values of $\beta$ is such that $\beta \neq (1 - 2\alpha)$ and consequently the denominator of Eq.~(\ref{Eqn: m relation on alpha and beta}) is never zero for the power law solution.

Note that it is also possible to express Eq.~(\ref{Eqn: alpha relation on m and n}) as a quadratic equation in $m$, which leads to the following roots for $m$.
\begin{equation} 
m \,=\, \frac{-1 + 2\al - n - 4\al n \,\pm \,\sqrt{W}}{2(2\al-1)},\label{Eqn: m quadratic equation soln}
\end{equation}
where 
\begin{equation} 
W\,=\, (1 - 2\al + n + 4
\al n)^2 - 4( 2\al -1)(2n - 4\al n + 6\al n^2).\label{Eqn: W}
\end{equation}
Recall that $m$ is the parameter in the power law solution $\phi(t) \propto t^m$.
For $m$ to be a real number, it follows from Eq.~(\ref{Eqn: m quadratic equation soln}) that this is possible only when $W \geq0$.
Similar to Eq.~(\ref{Eqn: S quadratic eqn}), we can express $W$ as
\begin{equation}
   W = \mathcal{Q}_2 n^{2} + \mathcal{Q}_1 n + \mathcal{Q}_0, \label{Eqn: W quadratic eqn}
\end{equation}
where 
\begin{eqnarray}
\mathcal{Q}_{_2}\, &=& \,  1 + 32\alpha - 32\alpha^2, \label{Eqn: Q1}\\
\mathcal{Q}_{_1}\, &=& \, 2\l(5 - 14\alpha + 8\alpha^2\r),\label{Eqn: Q2}\\
\mathcal{Q}_{_0}\, &=& \, (1 - 2\alpha)^2.\label{Eqn: Q3}
\end{eqnarray}
Since $\alpha$ is a positive integer greater than one, it is evident from Eq.~(\ref{Eqn: Q1}) that $\mathcal{Q}_{_2}$ is a negative number. 
This would mean that the $W$ described by Eq.~(\ref{Eqn: W quadratic eqn}) is a parabola on $n$ that opens up downwards. 
Consequently, to have $W \geq0$, it is required that $n$ must lie between the two roots of the quadratic equation $\mathcal{Q}_2 n^{2} + \mathcal{Q}_1 n^{2} + \mathcal{Q}_0 = 0$.
Therefore, a real value for $m$ in Eq.~(\ref{Eqn: m quadratic equation soln}) which requires $W \geq0$ restrict the value of $n$ to the following
\begin{equation} 
N \leq\; n \;\leq N',
\end{equation}
where,
\begin{eqnarray}
N \,=\,
\frac{8\alpha^{2} - 14\alpha + 5}{32\alpha^{2} - 32\alpha - 1}
- 2\sqrt{6}\,\sqrt{\frac{8\alpha^{4} - 20\alpha^{3} + 18\alpha^{2} - 7\alpha + 1}{\bigl(32\alpha^{2} - 32\alpha - 1\bigr)^{2}}},\label{Eqn: N}\\
N' \,=\,\frac{8\alpha^{2} - 14\alpha + 5}{32\alpha^{2} - 32\alpha - 1}
+ 2\sqrt{6}\,\sqrt{\frac{8\alpha^{4} - 20\alpha^{3} + 18\alpha^{2} - 7\alpha + 1}{\bigl(32\alpha^{2} - 32\alpha - 1\bigr)^{2}}}.\label{Eqn: N'}
\end{eqnarray}

\begin{figure}[t] 
  	\includegraphics[scale=0.5]{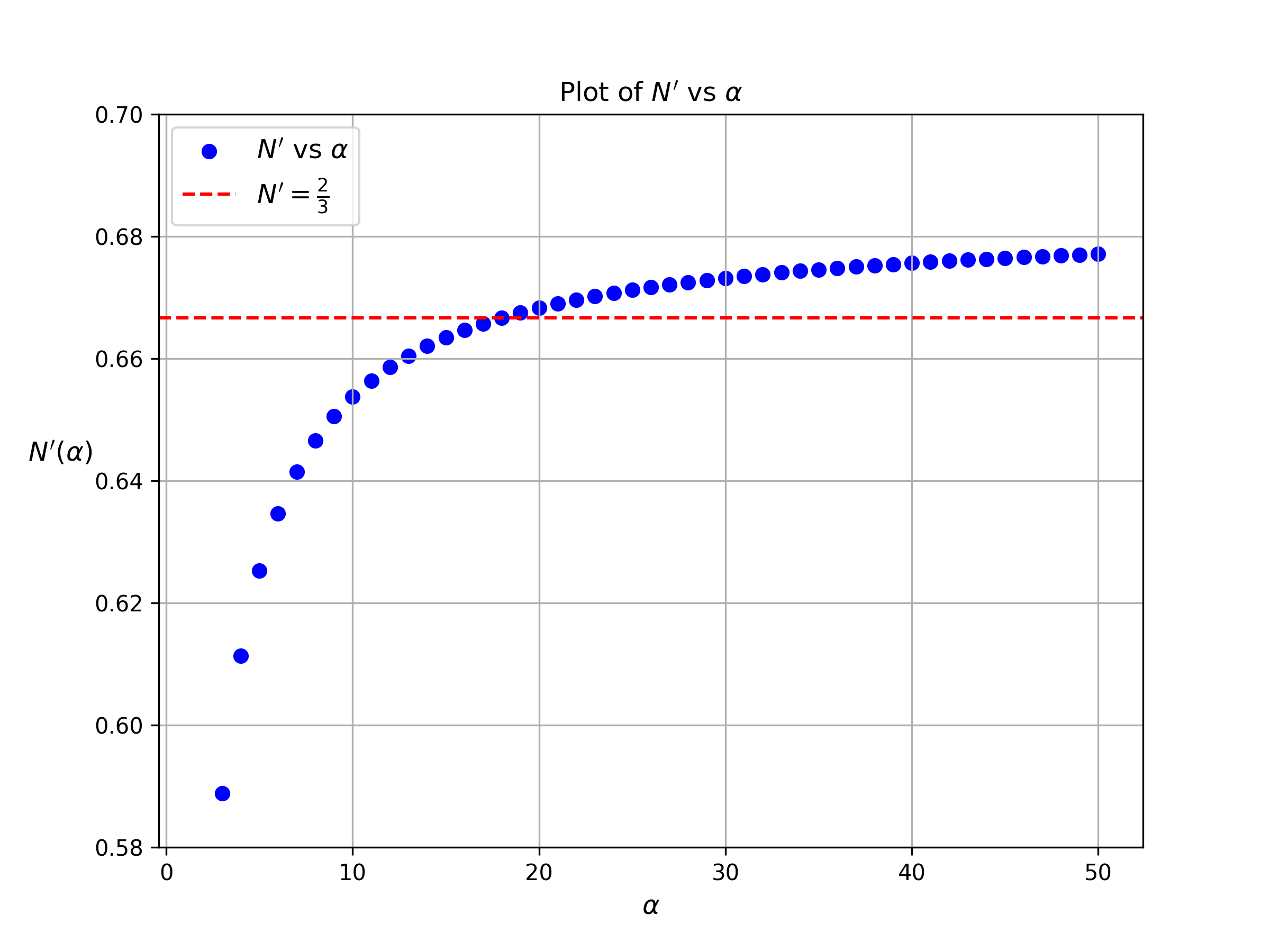}
  	\caption{Plot of $N'$ for different integer values of $\alpha$ from $\al = 2$ to $50$. Note that $N'$ as determined by Eq.~(\ref{Eqn: N'}) gives the upper bound on $n$ in the solution $a(t) \propto t^n$ for each value of $\alpha$. The blue dots represent the function $N'(\alpha)$ while the red dashed line indicates the constant value $N' = 2/3$. It is evident from this figure that only when $\alpha \geq18$ it is possible to have a solution $a(t) \propto t^{2/3}$.
 }
  	\label{Fig: N' vs alpha}
\end{figure}
For each integer value of $\alpha$ with $\alpha > 1$, Eq.~(\ref{Eqn: N'}) for $N'$ gives the maximum allowed value of $n$ in the power law solution $a(t) \propto t^{n}$.
Figure~(\ref{Fig: N' vs alpha}) shows the values of $N'$ for different integer values of $\alpha$ with $\alpha \, > \, 1$.
Note that, as $\alpha \rightarrow\ \infty$, it follows from Eq.~(\ref{Eqn: N'}) that $ N' \rightarrow\ (1+\sqrt{3})/4 \approx 0.683$.
Therefore, the power law expansion of the Universe with $a(t) \propto t^{n}$ is possible in the case of kinetic term driven expansion in the action~(\ref{Eqn: Action our new NC STT}) with $\el(X, \phi) = \lambda X^{\alpha}\phi^{\beta}$ but the maximum allowed value of $n$ is $0.683$.
This is one of the main result of this paper.

Evidently, with only the kinetic term $\lambda X^{\alpha}\phi^{\beta}$ in the action~(\ref{Eqn: Action our new NC STT}), it is not possible to have an accelerated expansion of the Universe which requires $n>1$ in the solution $a(t) \propto t^{n}$.
However, $n = 2/3$ is possible as it is less than the upper bound of $n = 0.683$.
In Einstein's GR, the expansion of the Universe during the cold dark matter dominated epoch is a power law expansion with $a(t) \propto t^{2/3}$.
It is possible to mimic this type of expansion in the model with action~(\ref{Eqn: Action our new NC STT}) when the kinetic term $\lambda X^{\alpha}\phi^{\beta}$ dominated the dynamics of the Universe.

With $n = 2/3$, Eq.~(\ref{Eqn: m quadratic equation soln}) becomes
\begin{equation} 
m = \frac{-5 - 2\al \pm \sqrt{73 - 76\al + 4\al ^2}}{ 12\al-6}. \label{Eqn: m eqn for n = 2/3} 
\end{equation}
Since $\alpha$ is a positive integer greater than one, to have a real value of $m$ from the above equation, it is required that 
\begin{equation}
  \alpha\; \geq \; 18.\label{Eqn: alpha ge 18}
\end{equation}
This point is also illustrated in the Figure~(\ref{Fig: N' vs alpha}), where it is shown that when $\alpha < 18$, the maximum allowed value $n$ denoted by $N'$ is less than $2/3$.

Hence, it is proved that the model described by action~(\ref{Eqn: Action our new NC STT}) with $\alpha\, \geq \, 18$ can lead to a solution $a(t) \propto t^{2/3}$ when $\lambda X^{\alpha}\phi^{\beta}$ dominates the dynamics of the Universe. 
This power law expansion is identical to the cold dark matted dominated expansion of the Universe based on Einstein's GR.
Hence, the kinetic term $\lambda X^{\alpha}\phi^{\beta}$ behaves as cold dark matter when the value of the parameter $\alpha\, \geq\, 18$.

Before we conclude this section, it is important to note that $a(t) \propto t^{2/3}$ is not possible in the case when the kinetic term dominates in the Brans-Dicke theory which corresponds to setting $\alpha = 1$ and $\beta = -1$ in the action~(\ref{Eqn: Action our new NC STT}). 
Following a similar analysis as presented in this section, the maximum allowed value of $n$ turns out to be $1/3$ assuming that the parameter $\lambda > 0$ in the action~(\ref{Eqn: Action our new NC STT}) along with $\alpha = 1$ and $\beta = -1$ which corresponds to the parameters for the Brans-Dicke theory.

Unlike the case of non-canonical scalar tensor theory, the value of $m$ and $n$ in the power law solution depends on the value of the parameter $\lambda$ in the Brans-Dicke case.
To see this, consider the action~(\ref{Eqn: Action our new NC STT}) without contributions from matter or the potential, and rescale the scalar field such that $\phi \to c_{_1}\tilde{\phi}$ for some dimensional constant $c_{_1}$. 
If we choose $c_{_1}$ such that $\lambda c_{_1}^{2\al+\be-1} = 1$, then the new action for the field $\tilde{\phi}$ does not depend on $\lambda$ and the two action differ by a multiplicative constant $c_{_1}$. 
However, for the particular case of Brans-Dicke theory, we do not have this freedom, since in this case $c_{_1}^{2\al+\be-1}=1$.
The implication of this is that, in the Brans-Dicke case, we cannot eliminate the dependence on $\lambda$.
With $\lambda>0$, one get the maximum value of $n = 1/3$.
This then rules out the possibility of $a(t) \propto t^{2/3}$ in the matter free expansion of the Universe in the Brans-Dicke theory without a potential term.
We therefore, in the next section, consider $\alpha\, \geq\, 18$ in the action~(\ref{Eqn: Action our new NC STT}) for the model to mimic the behavior of cold dark matter initially.

\section{$\Lambda$CDM like evolution of $a(t)$}\label{Sec: LCDM evolution}
We consider a matter free ($\el_m = 0$) spatially flat Universe based on action~(\ref{Eqn: Action our new NC STT}).
In the preceding section, we have shown that the kinetic term $\lambda X^{\alpha}\phi^{\beta}$ can behave as cold dark matter which leads to a solution $a(t) \propto t^{2/3}$.
Further, as mentioned in Sec.~(\ref{Sec: power law evolution}), a linear potential term in the action~(\ref{Eqn: Action our new NC STT}) can act as a cosmological constant term which is evident when one compares the two Friedmann equations~(\ref{Eqn: Friedman Eqn-1 Our GNC-STT}) and (\ref{Eqn: Friedman Eqn-2 Our GNC-STT}) with the corresponding equations  \emph{viz}..\ Eqs~(\ref{Eqn: Friedman Eqn-1 Lamda CDM}) and (\ref{Eqn: Friedman Eqn-2 Lambda CDM}) in the $\Lambda$CDM model based on Einstein's GR.
With a linear potential term and with $\el_m = 0$, the action~(\ref{Eqn: Action our new NC STT}) becomes
\begin{equation}
\mathcal{S} = \frac{1}{16\pi} \int \left[ -\phi R \,+\, \lambda X^{\alpha}\phi^{\beta}\,-\,V_{_0}\phi \right] \sqrt{-g}\,d^4 x.
\label{Eqn: Matterfree action our new NC STT}
\end{equation}

Our aim in this section is to determine whether the evolution of scale factor $a(t)$ in this model is identically the same as one gets in the $\Lambda$CDM model.
With a linear potential of the form $V(\phi) = V_{_0}\phi$ along with $\rho = p = 0$, the two Friedmann equations ~(\ref{Eqn: Friedman Eqn-1 Our GNC-STT}) and (\ref{Eqn: Friedman Eqn-2 Our GNC-STT}) and the field equation for $\phi$ given by Eq.~(\ref{Eqn: phi dot dot our GNC STT}) becomes
\begin{eqnarray}
\frac{\dot{a}^2}{a^2} &=&    - H\l(\frac{\dot \phi}{\phi}\r) \,+\, \l(\frac{2\alpha - 1}{6}\r)\lambda X^{\alpha}\phi^{\beta-1}\, +\, \frac{V_{_0}}{6},     \label{Eqn: Friedman Eqn-1 matterfree GNC-STT}\\
\frac{\ddot{a}}{a} &=&    - \frac{\ddot \phi}{2 \phi}\, - \,\frac{H \dot\phi}{2\phi}\, - \,\l(\frac{\alpha + 1}{6}\r)\lambda X^{\alpha}\phi^{\beta-1}\, +\, \frac{V_{_0}}{6},     
\label{Eqn: Friedman Eqn-2 Matterfree GNC-STT}\\
\ddot \phi &=& \frac{-3H\dot\phi\l[\alpha\lambda X^{\alpha-1}\phi^{\beta+1} + 3\r] + \l[\alpha +\beta - 2\alpha\beta -2\r]\lambda X^{\alpha}\phi^{\beta}+ V_{_0}\phi}{\alpha (2\alpha -1)\lambda X^{\alpha-1}\phi^{\beta+1} + 3}.\label{Eqn: phi dot dot matterfree GNC STT}
\end{eqnarray}

To numerically determine the solution $a(t)$ in this model, we first non-dimensionalise the above set of equations.
We consider the following dimensionless variables
\begin{eqnarray}
\tau &=&  H_{_i}t,  \label{Eqn: tau dimensionless}\\
\bar{a} &=&  \frac{a}{a_{_i}},  \label{Eqn: a bar}\\
\bar{\phi} &=&  \frac{\phi}{\phi_{_i}},  \label{Eqn: phi bar}
\end{eqnarray}
where $a_{_i}$ and $\phi_{_i}$ are the values of the scale factor and scalar field, respectively, at an initial time $t = t_{_i}$.
We consider the initial time as the matter dominated epoch with redshift $z \approx 1000$.
In Eq.~(\ref{Eqn: tau dimensionless}), $H_{_i}$ is the value of the Hubble parameter $H = \dot{a}/a$ at $t = t_{_i}$.
The variable $\tau$ can be interpreted as the dimensionless time variable.

In terms of the dimensionless variables $\bar{a}$ and $\bar{\phi}$, the two Friedmann equations~(\ref{Eqn: Friedman Eqn-1 matterfree GNC-STT}) and (\ref{Eqn: Friedman Eqn-2 Matterfree GNC-STT}) can be expressed as
\begin{eqnarray}
\l(\frac{\bar{a}'}{\bar{a}\,}\r)^{2} \,&=&\,    - \bar{H}\l(\frac{\bar{\phi}'}{\phi}\r) \,+\, \l(\frac{(2\alpha - 1)A}{6}\r)(\bar{\phi}')^{2\alpha}\bar{\phi}^{\beta-1}    \, +  \, U,  \label{Eqn: Friedman Eqn-1 dimensionless}\\
\frac{\bar{a}''}{\bar{a}} \,&=& \,   - \frac{1}{2}\l(\frac{ \bar{\phi}''}{\bar{\phi}\,}\r)\,-\,\l(\frac{\bar{H}\bar{\phi}'}{2\phi}\r)\,-\, \l(\frac{\alpha +1}{6} \r)A(\bar{\phi}')^{2\alpha}\bar{\phi}^{\beta-1}    \, +  \, U,  \label{Eqn: Friedman Eqn-2 dimensionless}
\end{eqnarray}
where the prime notation, like $\bar{\phi}'$, denotes a derivative with respect to the dimensionless time parameter $\tau$ and
\begin{eqnarray}
A \,&=&\,  \frac{\lambda H_{_i}^{2(\alpha-1)}\phi_{_i}^{2\alpha + \beta -1} }{2^{\alpha}},   \label{Eqn: A in terms of phi-i}\\
U \,&=& \, \frac{V_{_0}}{6H_{_i}^{2}}. \label{Eqn: U in terms of H-i}
\end{eqnarray}
Note that in these equations $\bar{H}$ is defined as 
\begin{equation}
\bar{H} \,=\, \frac{1}{\bar{a}}\l(\frac{\mathrm{d}\bar{a}}{\mathrm{d}\tau}\r).\label{Eqn: H bar}
\end{equation}

Similar to the above Friedmann equations in terms of dimensionless variables, the field equations for $\bar{\phi}$ take the form
\begin{equation}
\frac{\bar{\phi}''}{\bar{\phi}\,} \,=\, \frac{\mathcal{N}_{T1} \,+ \, \mathcal{N}_{T2}}{\mathcal{D}_{T}},\label{Eqn: phi prime prime dimensionless}
\end{equation}
where 
\begin{eqnarray}
\mathcal{N}_{T1} \,&=& \, -\l(\frac{3\bar{H}\bar{\phi}'}{\phi}\r)\l[2\alpha A (\bar{\phi}')^{2(\alpha-1)}\bar{\phi}^{\beta+1}\,+\,3\r],\label{Eqn: NT1}\\
\mathcal{N}_{T2} \,&=& \, \l(\alpha + \beta - 2\alpha\beta -2\r)A(\bar{\phi}')^{2\alpha}\bar{\phi}^{\beta-1}    \, +  \, 6U, \label{Eqn: NT2}\\
\mathcal{D}_{T} \,&=& \, 2\alpha (2\alpha -1) A (\bar{\phi}')^{2(\alpha-1)}\bar{\phi}^{\beta+1}\,+\,3.\label{Eqn: D}
\end{eqnarray}

As shown in the preceding section, the term $\lambda X^{\alpha}\phi^{\beta}$ can lead to $a(t) \propto t^{2/3}$ and $\phi(t) \propto t^{m}$ for $\alpha \geq 18$.
Therefore, we consider $\alpha \geq 18$ and choose the value of the parameter $\beta$ such that initially during the kinetic term $\lambda X^{\alpha}\phi^{\beta}$ dominated epoch $\bar{a}(\tau)$ and $\bar{\phi}(\tau)$ evolves as
\begin{eqnarray}
\bar{a}(\tau)\,&\approx& \, \l(\frac{\tau}{\tau_{i}}\r)^{\frac{2}{3}},\label{Eqn: a(tau) initially}\\
\bar{\phi}(\tau) \,&\approx& \, \l(\frac{\tau}{\tau_{i}}\r)^{m},\label{Eqn: phi(tau) initially}
\end{eqnarray}
where $m$ is fixed for a given value of $\alpha$ using Eq.~(\ref{Eqn: m eqn for n = 2/3}). 
Although Eq.~(\ref{Eqn: m eqn for n = 2/3}) gives two roots of $m$, we consider the one with the $+$ sign among the $\pm$ signs in Eq.~(\ref{Eqn: m eqn for n = 2/3}). However, it is important to emphasize that the other root of $m$ also leads to similar results.

Once the value of $m$ is determined, the parameter $\beta$ is then fixed using Eq.~(\ref{Eqn: beta relation}).
Note that Eqs.~(\ref{Eqn: a(tau) initially}) and (\ref{Eqn: phi(tau) initially}) are used only to fix the parameter $\beta$ and to determine the value of $m$ for each value of $\alpha$ and also to fix the initial condition for $\bar{a}'$ and $\bar{\phi}'$.
Whether or not $\bar{a}(\tau)$ and $\bar{\phi}(\tau)$ evolve as power law form initially during the kinetic term dominated epoch will be decided by the numerical solutions of the Eqs.~(\ref{Eqn: Friedman Eqn-1 dimensionless}), (\ref{Eqn: Friedman Eqn-2 dimensionless}) and (\ref{Eqn: phi prime prime dimensionless}). 
However, it is important to note that from the analysis of the preceding section, it is evident that with an appropriate choice of parameters $\alpha$ and $\beta$, the model with action~(\ref{Eqn: Matterfree action our new NC STT}) does lead to a power law solution of the form given by Eqs.~(\ref{Eqn: a(tau) initially}) and (\ref{Eqn: phi(tau) initially}) when the potential term is subdominant.

From the definition of the dimensionless variables $\bar{a}$ and $\bar{\phi}$ and using the assumed initial form of evolution of $\bar{a}(\tau)$ and $\bar{\phi}(\tau)$, it follows that the initial conditions for solving the differential equations Eqs.~(\ref{Eqn: Friedman Eqn-1 dimensionless}), (\ref{Eqn: Friedman Eqn-2 dimensionless}) and (\ref{Eqn: phi prime prime dimensionless}) are given by
\begin{equation}
 \bar{a}(\tau_{i})\, = \,\bar{\phi}(\tau_{i}) \,= \,\bar{a}'(\tau_{i})\, =\, 1~~~~~\mathrm{and}~~~~~\bar{\phi}'(\tau_{i}) = \frac{3m}{2},\label{Eqn: initial conditions}
\end{equation}
where $\tau_{i}$ is the initial value of the dimensionless time $\tau$ defined in Eq.~(\ref{Eqn: tau dimensionless}).
From Eq.~(\ref{Eqn: a(tau) initially}), it follows that $\tau_{i} = 2/3$.
Note that we are considering the initial time as the matter dominated epoch with redshift $z \approx 1000$.
Since $1+z = a_{_0}/a$, where $a_{_0}$ is the value of the scale factor at the present epoch, it follows from Eq.~(\ref{Eqn: a bar}) that at the present epoch $\bar{a} = 1000$. 
We therefore numerically solve differential equations for $\bar{a}(\tau)$ and  $\bar{\phi}(\tau)$ from $\tau = \tau_{i} = 2/3$ to $\tau = \tau_{0}$, where $\tau_{0}$ is the value of $\tau$ at the present epoch when $\bar{a}(\tau_{0}) = 1000$.

Note that Eq.~(\ref{Eqn: Friedman Eqn-1 dimensionless}) along with the initial conditions~(\ref{Eqn: initial conditions})  implies that the constant $A$ defined in Eq.~(\ref{Eqn: A in terms of phi-i}) takes the following value
\begin{equation}
A \,=\, \frac{(3\times2^{2\alpha})(2 + 3m - U)}{(2\alpha - 1)(3m)^{2\alpha}}.   \label{Eqn: A numberic value}
\end{equation}
The value of $U$ given by Eq.~(\ref{Eqn: U in terms of H-i}) can be fixed in the following way.
On comparing Eq.~(\ref{Eqn: Friedman Eqn-1 Lamda CDM}) with Eq.~(\ref{Eqn: Friedman Eqn-1 matterfree GNC-STT}), it is clear that $V_{_0} = 2\Lambda$ where $\Lambda$ is the cosmological constant in Einstein's GR whose value can be expressed as
\begin{equation}
\Lambda \,=\, 3H_{_0}^{2}(1- \Omega_{_{m0}}),   \label{Eqn: Lambda}
\end{equation}
where $\Omega_{_{m0}} = (8 \pi G \rho_{_{m0}})/(3H_{0}^{2})$ is the density parameter of the cold dark matter at the present epoch.
We take $\Omega_{_{m0}} = 0.31$ in this paper. 
In Eq.~(\ref{Eqn: Lambda}), $H_{0}$ is the Hubble constant at the present epoch. 
From Eq.~(\ref{Eqn: Friedman Eqn-1 Lamda CDM}), it follows that initially during matter dominated epoch $H_{i}^{2} = H_{0}^{2}\Omega_{_{m0}}a_{_i}^{-3}$.
Therefore, Eqs.~(\ref{Eqn: U in terms of H-i}) and (\ref{Eqn: Lambda}) implies that
\begin{equation}
U \,=\, \l(\frac{1- \Omega_{_{m0}}}{\Omega_{_{m0}}}\r)a_{_i}^{-3}.\label{Eqn: U numberic value}
\end{equation}
We consider the initial matter dominated epoch as an epoch at redshift $z = 1000$ which gives $a_{_i} \approx 10^{-3}$.

It is important to note that not all three equations  \emph{viz}.. Eqs.~(\ref{Eqn: Friedman Eqn-1 dimensionless}), (\ref{Eqn: Friedman Eqn-2 dimensionless}) and (\ref{Eqn: phi prime prime dimensionless}) are independent.
We, therefore, use the two second order differentials equations,  namely, Eqs.~(\ref{Eqn: Friedman Eqn-2 dimensionless}) and (\ref{Eqn: phi prime prime dimensionless}) to numerically determine the solution $\bar{a}(\tau)$ and   $\bar{\phi}(\tau)$ and use Eq.~(\ref{Eqn: Friedman Eqn-1 dimensionless}) as constraint equation to verify the consistency of the numerical solution.
For this, we define the following quantity $f(\tau)$ defined as 
\begin{equation}
f(\tau) \,=\,    \l(\frac{\bar{a}'}{\bar{a}\,}\r)^{2}\l[- \bar{H}\l(\frac{\bar{\phi}'}{\phi}\r) \,+\, \l(\frac{(2\alpha - 1)A}{6}\r)(\bar{\phi}')^{2\alpha}\bar{\phi}^{\beta-1}    \, +  \, U\r]^{-1}.  \label{Eqn: f(tau)}
\end{equation}

\begin{figure}[t] 
\includegraphics[scale=0.40]{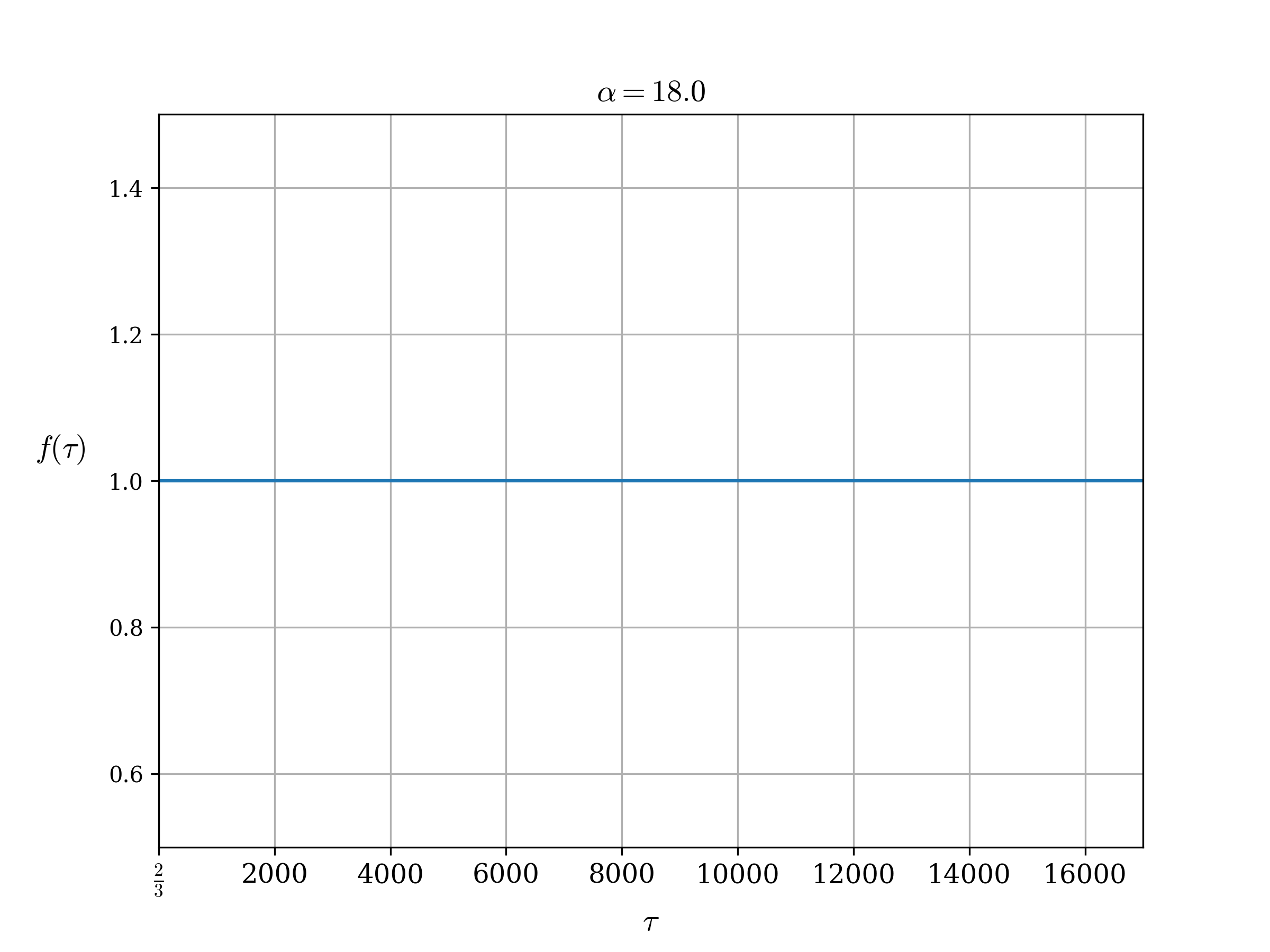}
\includegraphics[scale=0.40]{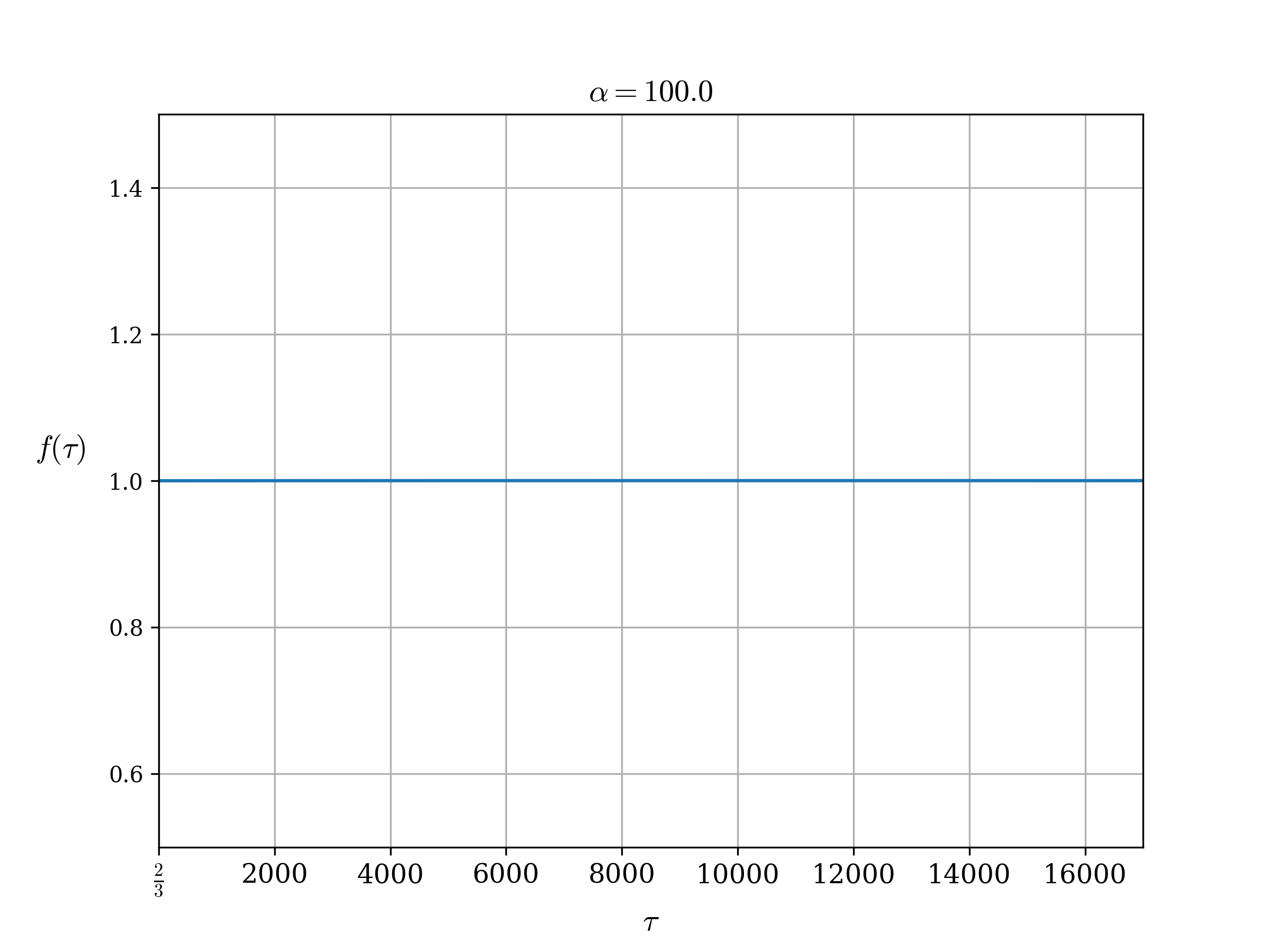}
\caption{Plot of $f(\tau)$ as a function of $\tau$ for $\alpha = 18$ and  $\alpha = 100$ where $f(\tau)$ is defined in Eq.~(\ref{Eqn: f(tau)}).  
 Evidently $f(\tau) = 1$ which confirms that the solutions  $\bar{a}(\tau)$ and   $\bar{\phi}(\tau)$ obtained from Eqs.~(\ref{Eqn: Friedman Eqn-2 dimensionless}) and (\ref{Eqn: phi prime prime dimensionless}) are consistent with Eq.~(\ref{Eqn: Friedman Eqn-1 dimensionless}).}
\label{Fig: f(tau)}
\end{figure}

After solving Eqs.~(\ref{Eqn: Friedman Eqn-2 dimensionless}) and (\ref{Eqn: phi prime prime dimensionless}) with the initial conditions given by Eq.~(\ref{Eqn: initial conditions}), we plot $f(\tau)$ as a function of $\tau$.
This plot is shown in Fig.~(\ref{Fig: f(tau)}). 
Clearly, $f(\tau) = 1$, thereby confirming that the solution $\bar{a}(\tau)$ and   $\bar{\phi}(\tau)$ obtained from Eqs.~(\ref{Eqn: Friedman Eqn-2 dimensionless}) and (\ref{Eqn: phi prime prime dimensionless})  are consistent with Eq.~(\ref{Eqn: Friedman Eqn-1 dimensionless}).

\subsection{Comparing with $a(t)$ in GR-$\Lambda$CDM model}
\begin{figure}[t] 
\includegraphics[scale=0.40]{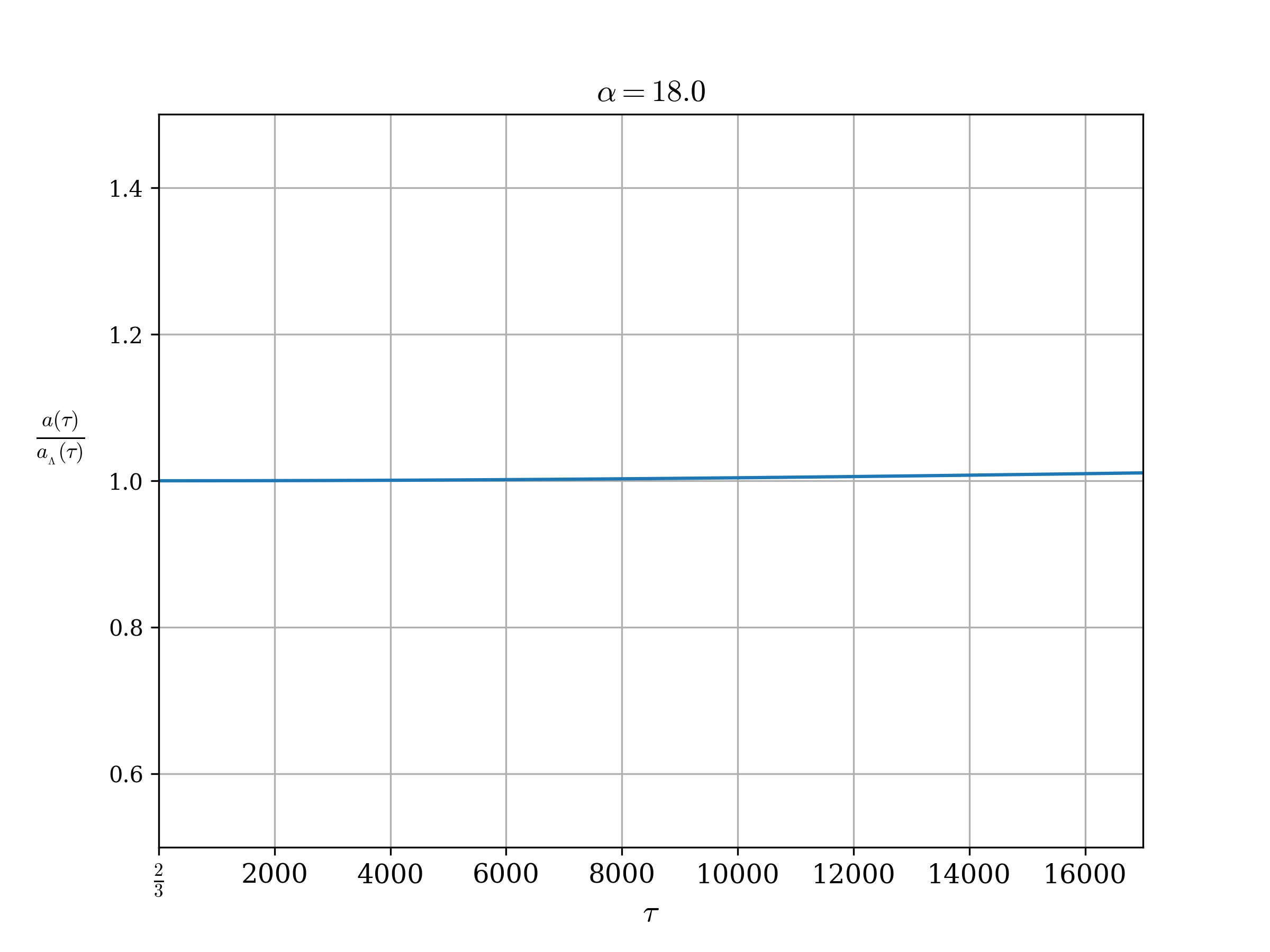}
\includegraphics[scale=0.40]{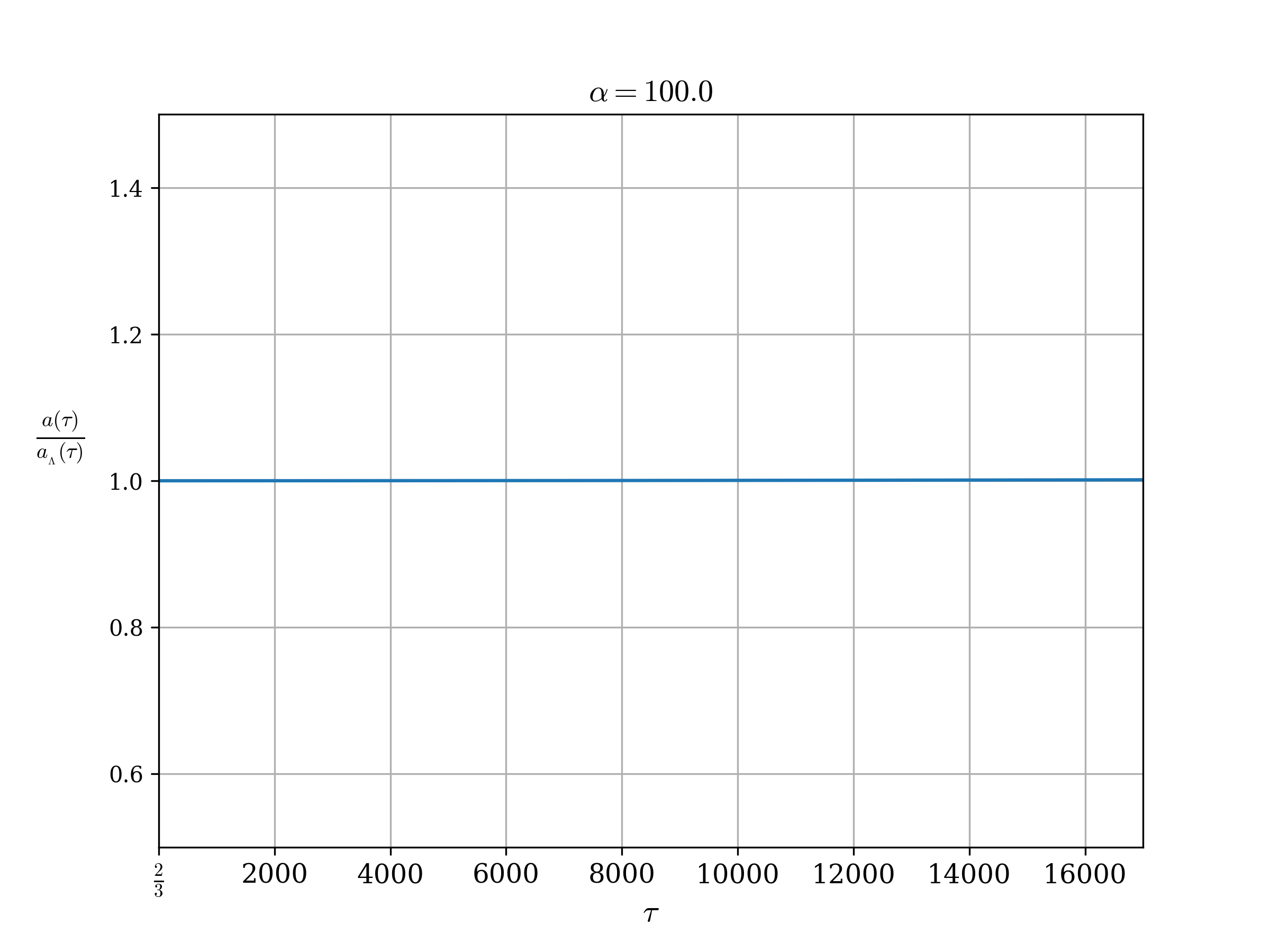}
\caption{Plot of the ratio $a(\tau)/a_{_\Lambda}(\tau)$ as a function of $\tau$ for $\alpha = 18$ and  $\alpha = 100$.
These plots clearly illustrates that $a(\tau) \approx a_{_\Lambda}(\tau)$ for $\alpha \geq 18$. The larger the value of  $\alpha$ closer is $a(\tau)$ to $a_{_\Lambda}(\tau)$. 
Both the solutions are observationally indistinguishable for $\alpha = 100$.}
\label{Fig: ratio a(tau)/a_L(tau)}
\end{figure}

Solving Eqs.~(\ref{Eqn: Friedman Eqn-2 dimensionless}) and (\ref{Eqn: phi prime prime dimensionless}) with the initial conditions~(\ref{Eqn: initial conditions}) gives $\bar{a}(\tau)$ and $\bar{\phi}(\tau)$.
We compare the solution $\bar{a}(\tau)$ with the corresponding solution for scale factor in the $\Lambda$CDM model based on Einstein's GR for which Eq.~(\ref{Eqn: Friedman Eqn-1 Lamda CDM}) leads to the following solution 
\begin{equation}
a_{_\Lambda}(\tau) \,=\, \l(\frac{\Omega_{_{m0}}}{1 - \Omega_{_{m0}}}\r)^{1/3}\l[\sinh\l(\sqrt{\frac{a_{_i}^{3}(1-\Omega_{_{m0}})}{\Omega_{_{m0}}}}\,\l(\frac{3\tau}{2}\r) \r)\r]^{\frac{2}{3}}.
\label{Eqn: a(tau) LCDM}
\end{equation}
The subscript $\Lambda$ in $a_{_\Lambda}(\tau)$ is for distinguishing this solution in $\Lambda$CDM model with the one in the model considered in this paper. 
In Fig.~(\ref{Fig: ratio a(tau)/a_L(tau)}) we plot the ratio $a(\tau)/a_{_\Lambda}(\tau)$ as a function of $\tau$ for two different values of $\alpha$.
It is evident from this figure that the scale factor $a(\tau)$ in this model with $\alpha \geq 18$ evolves just like the same in the GR based $\Lambda$CDM model.
The larger the value of $\alpha$, the closer  the solution for $a(\tau)$ is to $a_{_\Lambda}(\tau)$.

\begin{figure}[t] 
\includegraphics[scale=0.40]{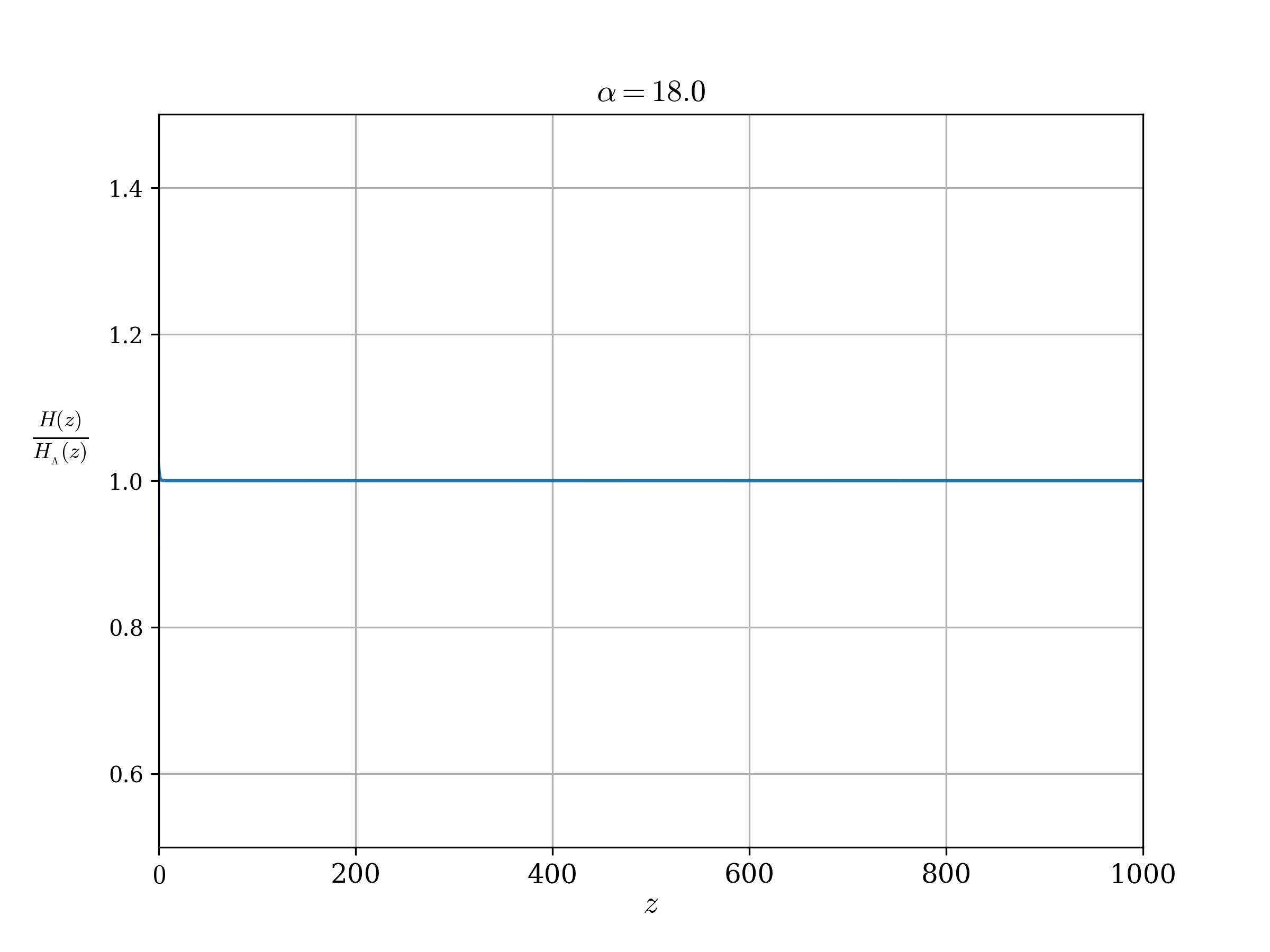}
\includegraphics[scale=0.40]{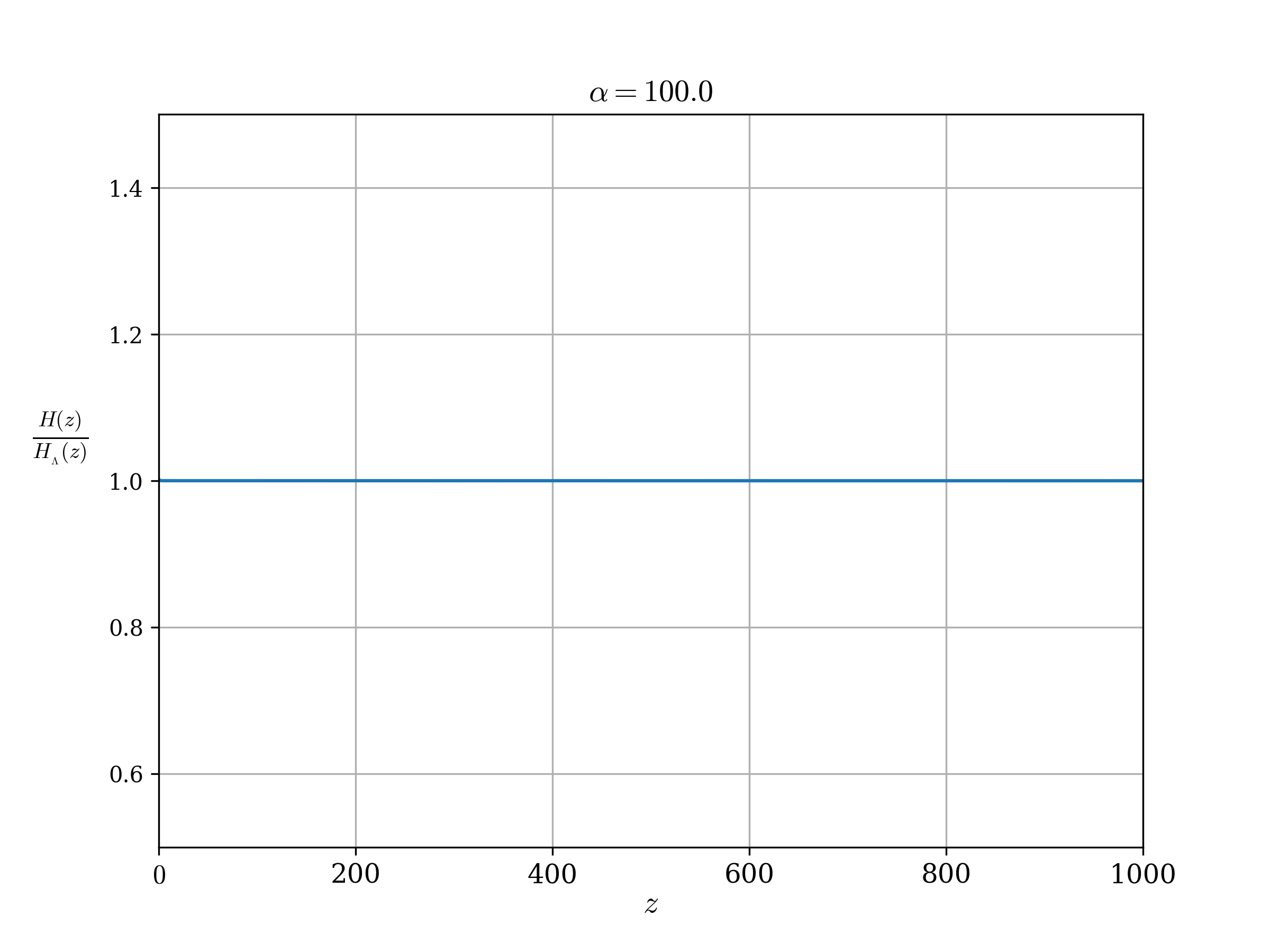}
\caption{Plot of the ratio $H(z)/H_{_\Lambda}(z)$ as a function of redshift $z$ for $\alpha = 18$ and  $\alpha = 100$. 
Here $H(z)$ is the Hubble parameter as a function of redshift $z$ in the model considered in this paper whereas $H_{_\Lambda}(z)$ is corresponding function in the GR- based  $\Lambda$CDM model.}
\label{Fig: ratio H(z)/H_L(z)}
\end{figure}

Note that in Fig.~(\ref{Fig: ratio a(tau)/a_L(tau)}), we plot the  ratio $a(\tau)/a_{_\Lambda}(\tau)$ from $\tau = \tau_{_i} = 2/3$ to $\tau = \tau_{_0}$ here $\tau_{_0}$ is the value of $\tau$ at the present epoch at which $a_{_\Lambda}(\tau_{_0}) = 1$ and this gives
\begin{equation}
\tau_{_0}\,=\, \frac{2}{3}\l(\sqrt{\frac{\Omega_{_{m0}}a_{_i}^{-3}}{1-\Omega_{_{m0}}}}\r)\sinh^{-1} \l(\sqrt{\frac{1-\Omega_{_{m0}}}{\Omega_{_{m0}}}}\,\r).
\label{Eqn: tau-0}
\end{equation}
With $a_{_i} = 10^{-3}$ and $\Omega_{_{m0}} = 0.31$, the above equation implies that $\tau_{_0} \approx 16820$. 
We have therefore plotted the ratio $a(\tau)/a_{_\Lambda}(\tau)$ in Fig.~(\ref{Fig: ratio a(tau)/a_L(tau)}) from $\tau = 2/3$ to $17000$.

Similar to the ratio $a(\tau)/a_{_\Lambda}(\tau)$, in Fig.~(\ref{Fig: ratio H(z)/H_L(z)}), we have plotted the ratio $H(z)/H_{_\Lambda}(z)$ as a function of redshift $z$,
 where $H(z)$ is the value of the Hubble parameter $H = \dot{a}/a$ for each $z$ in the model considered in this paper.
The corresponding $H_{_\Lambda}(z)$ is for the GR based $\Lambda$CDM model.
It is evident from Fig.~(\ref{Fig: ratio H(z)/H_L(z)}) that $H(z) \approx H_{_\Lambda}(z)$, which once again illustrates the fact that the model described in this paper mimics the GR-based $\Lambda$CDM model.

The deceleration parameter $q$ is defined as 
\begin{equation}
q \,=\, -\frac{\ddot{a}}{a H^{2}}.   \label{Eqn: deceleration parameter}
\end{equation}

\begin{figure}[t] 
\includegraphics[scale=0.40]{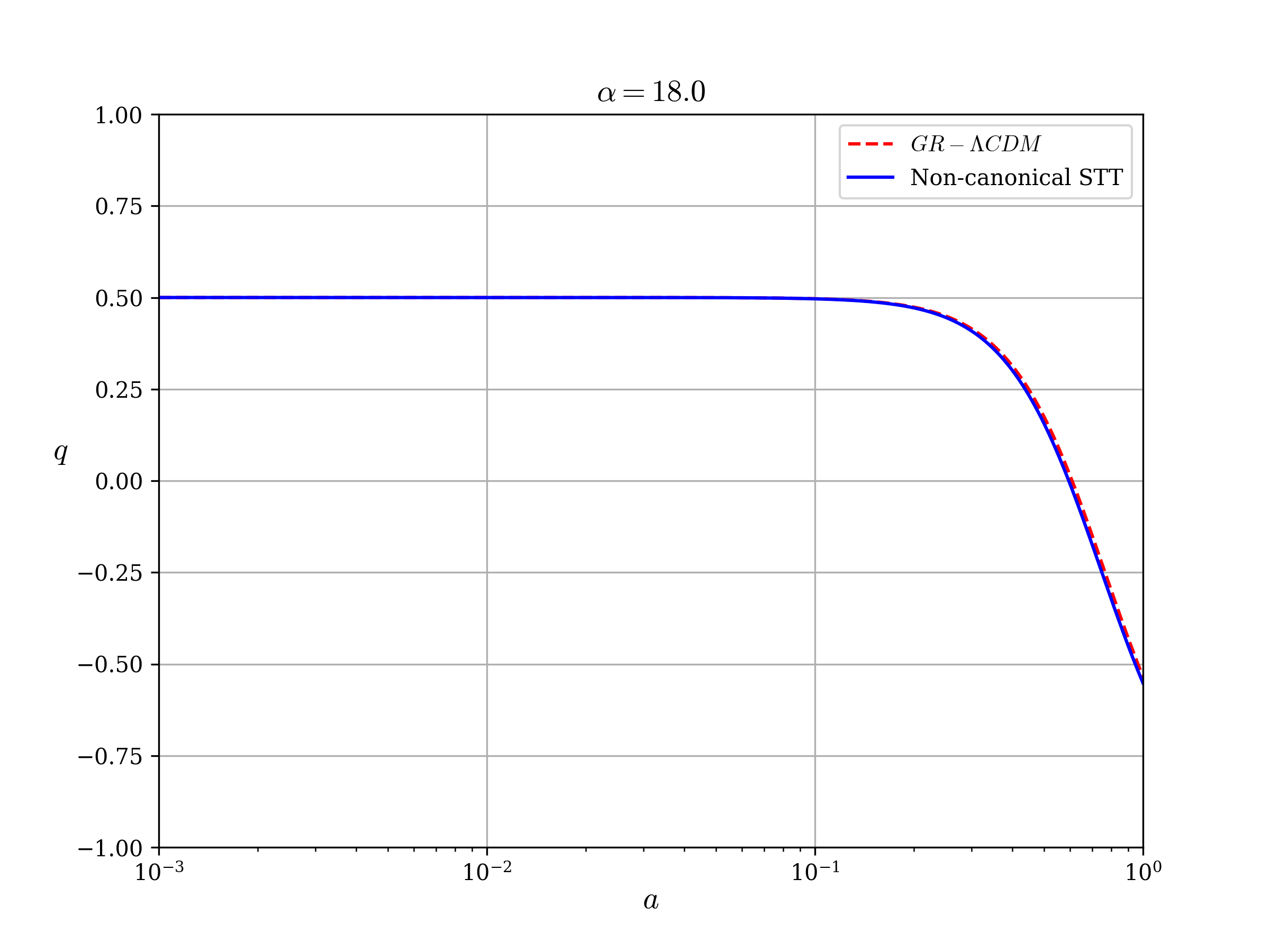}
\includegraphics[scale=0.40]{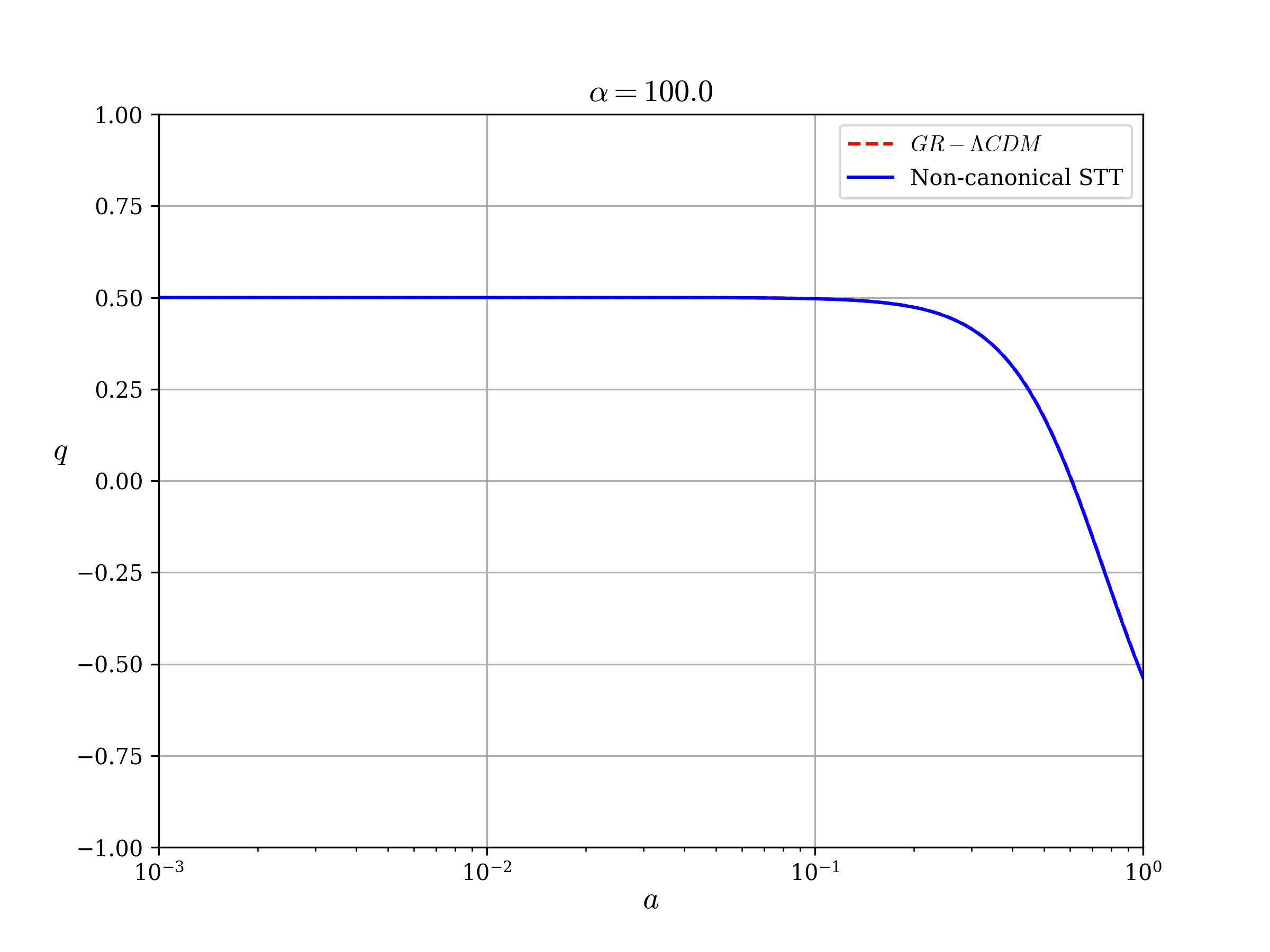}
\caption{Plot of the deceleration parameter $q$ as a function of scale factor $a$ for $\alpha = 18$ and  $\alpha = 100$.
The blue line shows the $q$ as a function of $a$ in the non-canonical scalar-tensor model considered in this paper, while the red dashed line gives $q$ for the GR based $\Lambda$CDM model.
The two curves nearly overlaps for $\alpha = 18$ while for $\alpha = 100$ the two curves are virtually indistinguishable.}
\label{Fig: deceleration parameter q}
\end{figure}

In the cold dark matter dominated epoch in the GR based $\Lambda$CDM model, it turns out that initially $q = 0.5$.
One gets the same value of $q = 0.5$ in the kinetic term $\lambda X^\alpha \phi^\beta$ dominated epoch in the model with action~(\ref{Eqn: Matterfree action our new NC STT}).
In Fig.~(\ref{Fig: deceleration parameter q}), we plot $q$ as a function of scale factor for two different value of $\alpha$.
For comparison, the deceleration parameter $q$ that one gets in the GR based $\Lambda$CDM model is also plotted in the same figure, which illustrates the fact that the evolution of $q$ in both models is virtually indistinguishable.

\subsection{The Equivalent Dark Matter and Dark Energy }
The two Friedmann equations  \emph{viz}. Eqs.~(\ref{Eqn: Friedman Eqn-1 matterfree GNC-STT}) and (\ref{Eqn: Friedman Eqn-2 Matterfree GNC-STT}), can be expressed in the same way as one gets in the Einstein's GR.
This leads to the following equations
\begin{eqnarray}
\frac{\dot{a}^2}{a^2}\, &=&\,  \l(\frac{8\pi G }{3}\r)\Big[\rho_{_{DM}}\, + \,  \rho_{_{DE}}\Big],   \label{Eqn: Friedman Eqn-1 NTT eqlnt GR}\\
\frac{\ddot{a}}{a}\,&=&\,    -\l(\frac{4 \pi G}{3}\r)\Big[\l(\rho_{_{DM}} + 3p_{_{DM}}\r)\, + \,  \l(\rho_{_{DE}}+ 3p_{_{DE}}\r)\Big],
\label{Eqn: Friedman Eqn-2 NTT eqlnt GR}
\end{eqnarray}
where 
\begin{align}
\rho_{_{\mathrm{DM}}} &= \left( \frac{3}{8\pi G} \right)
\left[
    -H \left( \frac{\dot{\phi}}{\phi} \right)
    + \frac{(2\alpha - 1)}{6} \lambda X^{\alpha} \phi^{\beta - 1}
\right],
\label{Eqn:rho_DM} \\[6pt]
p_{_{\mathrm{DM}}} &= \left( \frac{3}{8\pi G} \right)
\left[
    \frac{\ddot{\phi}}{3\phi}
    + \frac{\lambda X^{\alpha} \phi^{\beta - 1}}{6}
    + \frac{2H \dot{\phi}}{3\phi}
\right],
\label{Eqn:p_DM} \\[6pt]
\rho_{_{\mathrm{DE}}} &= \frac{V_{_0}}{16\pi G},
\label{Eqn:rho_DE} \\[6pt]
p_{_{\mathrm{DE}}} &= -\rho_{_{\mathrm{DE}}}.
\label{Eqn:p_DE}
\end{align}

\begin{figure}[t] 
\includegraphics[scale=0.40]{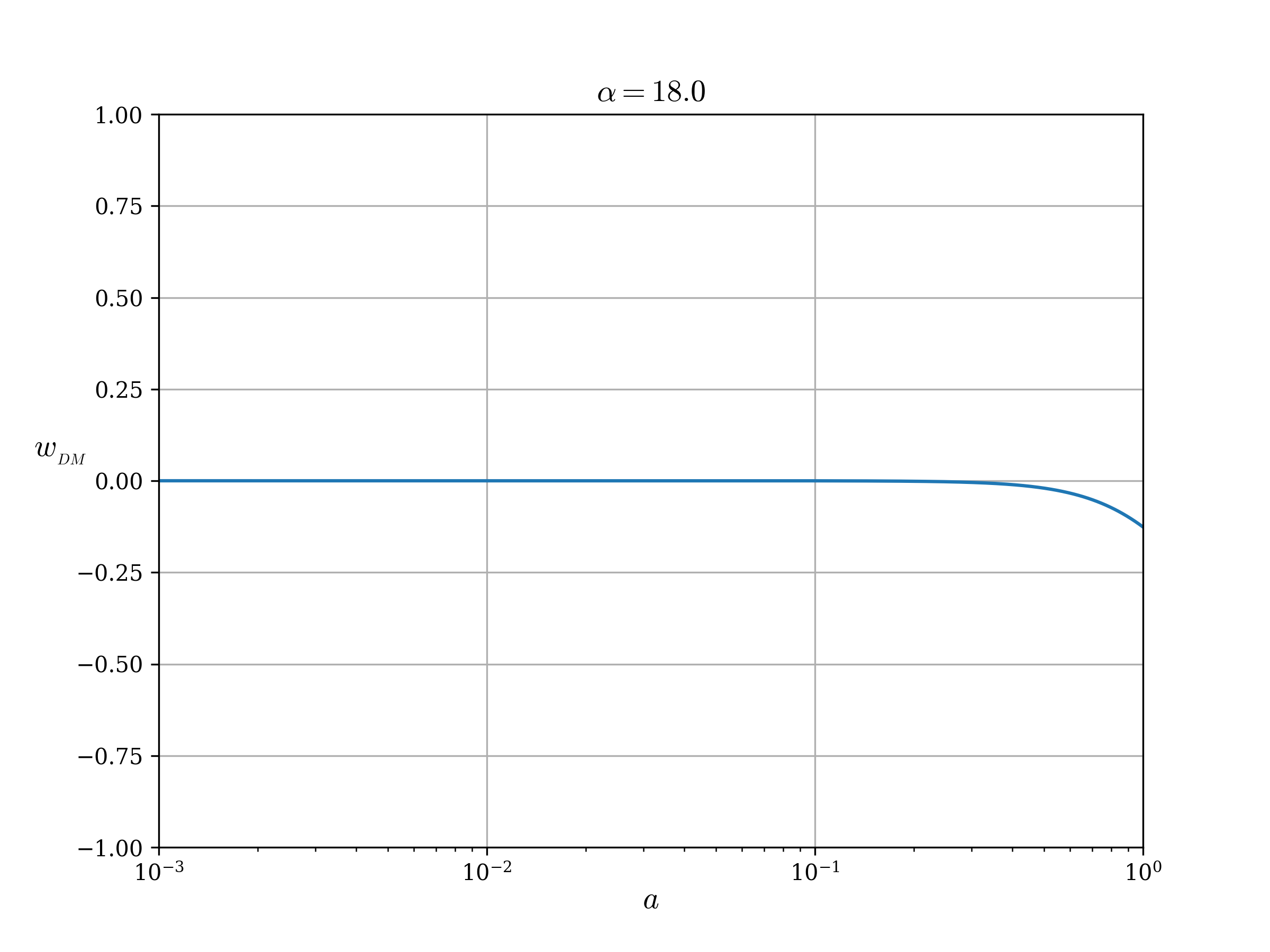}
\includegraphics[scale=0.40]{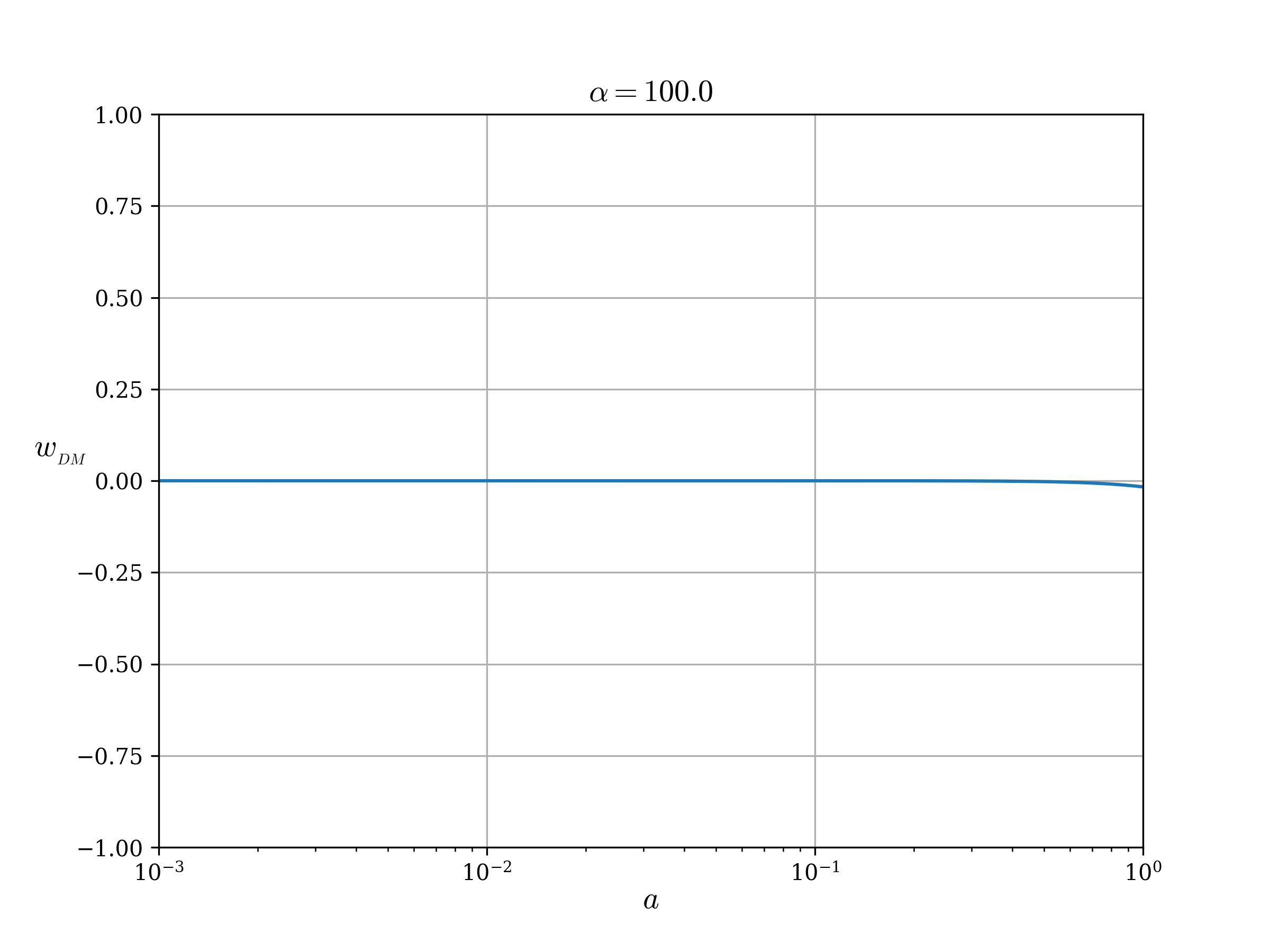}
\includegraphics[scale=0.40]{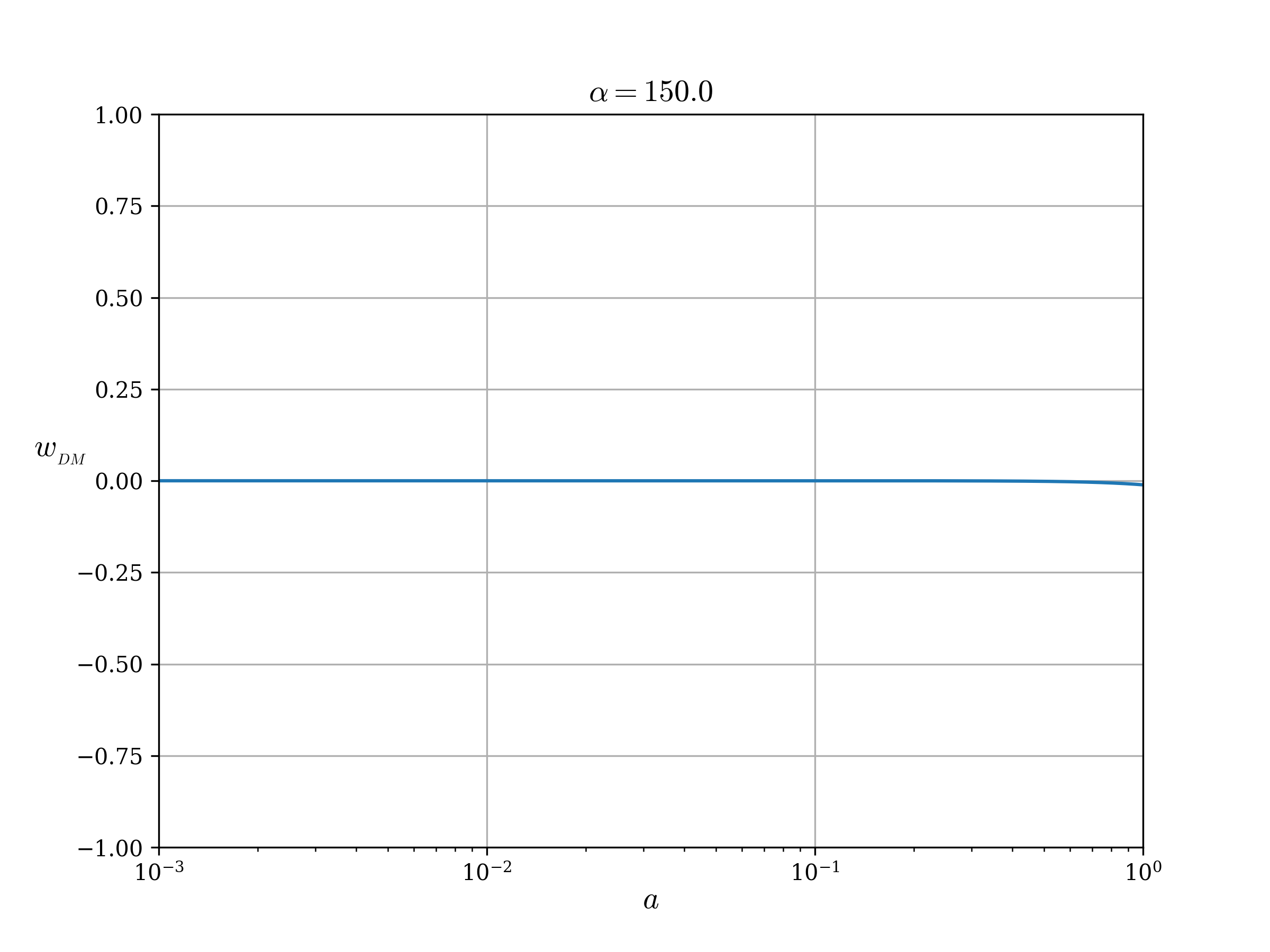}
\includegraphics[scale=0.40]{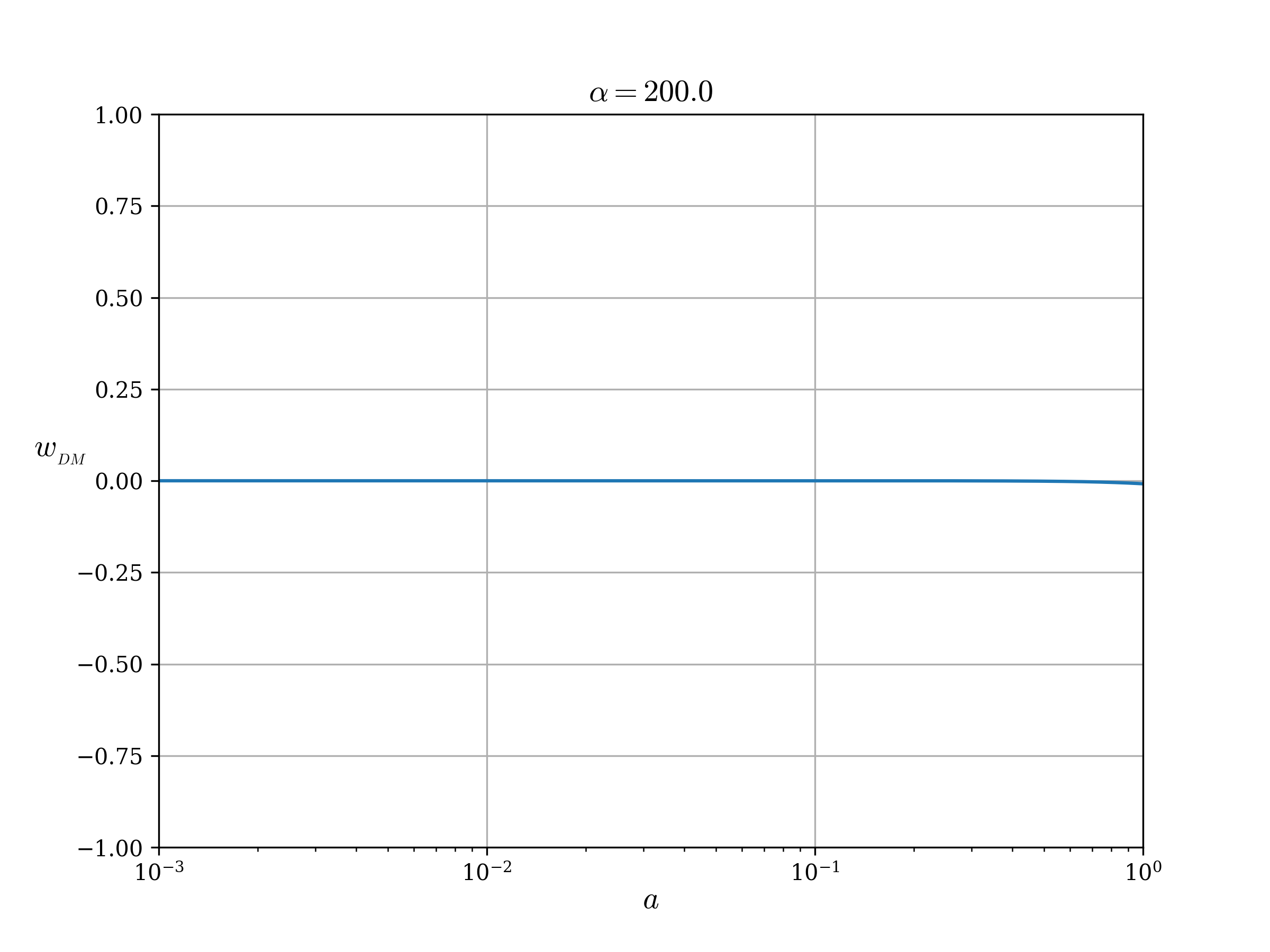}
\caption{Plot of the equation of state parameter $w_{_{DM}}$ of the dark matter part as a function of scale factor $a$ for $\alpha = 18, 100, 150$ and $200$.
For large value of $\alpha$, evidently $w_{_{DM}} \approx 0$.}
\label{Fig: wDM)}
\end{figure} 

Clearly, from Eq.~(\ref{Eqn:p_DE}), the equation of state parameter of the dark energy part given by $w_{_{DE}} = p_{_{DE}}/\rho_{_{DE}} = -1$ which is identically the same as one gets for the cosmological constant. 
For the dark matter part, $w_{_{DM}} = p_{_{DM}}/\rho_{_{DM}}$.
As illustrated in the preceding section, in the absence of the potential term in the action~(\ref{Eqn: Matterfree action our new NC STT}), the kinetic term $\lambda X^{\alpha}\phi^{\beta}$ with $\alpha \geq 18$ leads to a solution $a(t) \propto t^{2/3}$.
This would then imply that $w_{_{DM}} = 0$.
This will also be the case when the potential term $V_{_0}\phi$ term is there but it is subdominant.
However, when the potential term starts to dominate, it is necessary to verify whether $w_{_{DM}}$ still remains zero.
We therefore plot $w_{_{DM}} = p_{_{DE}}/\rho_{_{DE}}$ using the solutions $\bar{a}(\tau)$ and   $\bar{\phi}(\tau)$ obtained from the two second order differentials equations,  namely, Eqs.~(\ref{Eqn: Friedman Eqn-2 dimensionless}) and (\ref{Eqn: phi prime prime dimensionless}).
This plot is shown in Fig.~(\ref{Fig: wDM)}).
It is evident from these plots that, for large values of $\alpha$, $w_{_{DM}} \approx 0$ even when the potential term starts to dominate.
This confirms that the dark matter part behaves like a pressureless fluid with $p_{_{DM}} \approx 0$ for large values of $\alpha$.

\begin{figure}[ht] 
\includegraphics[scale=0.45]{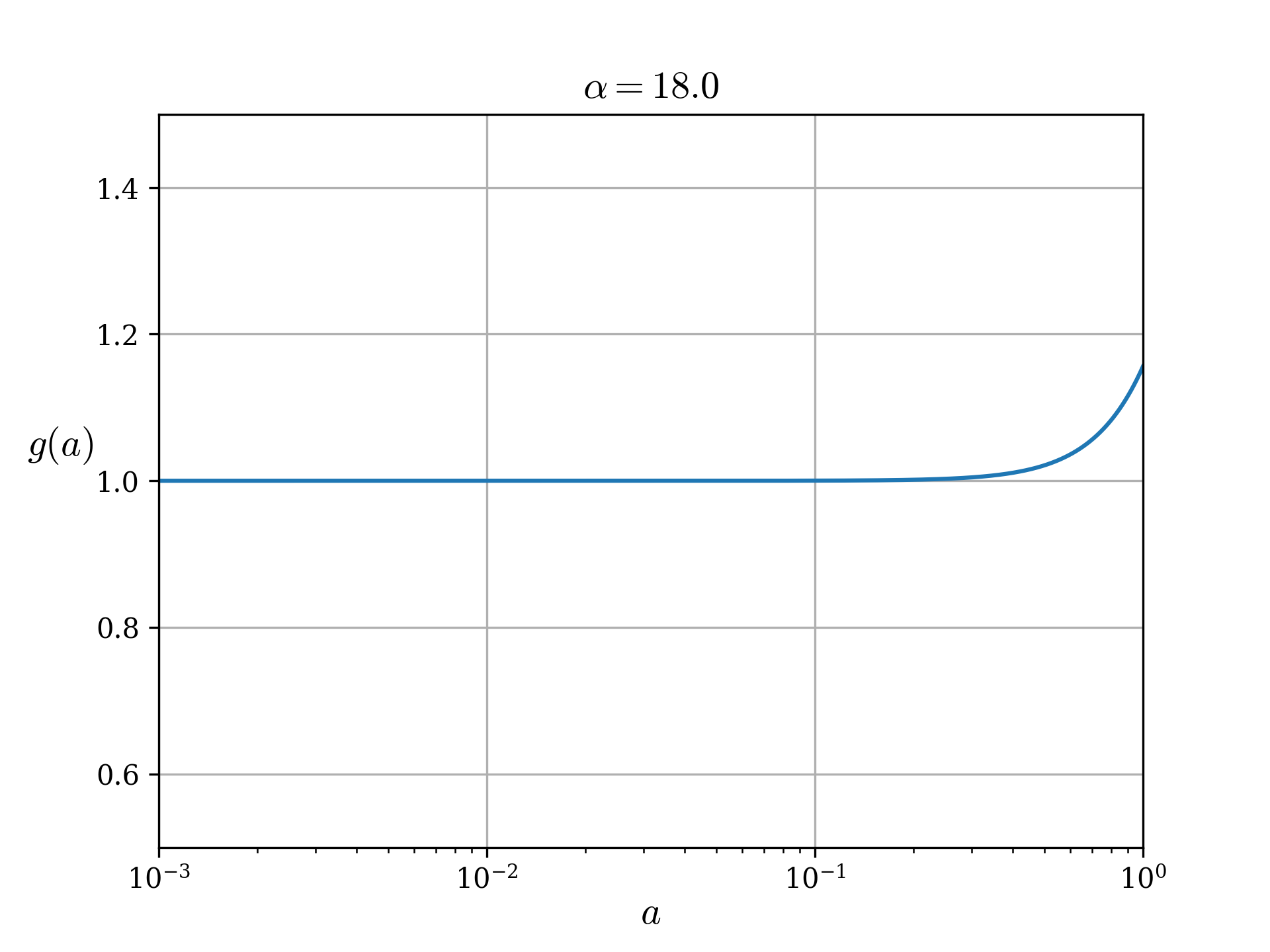}
\includegraphics[scale=0.45]{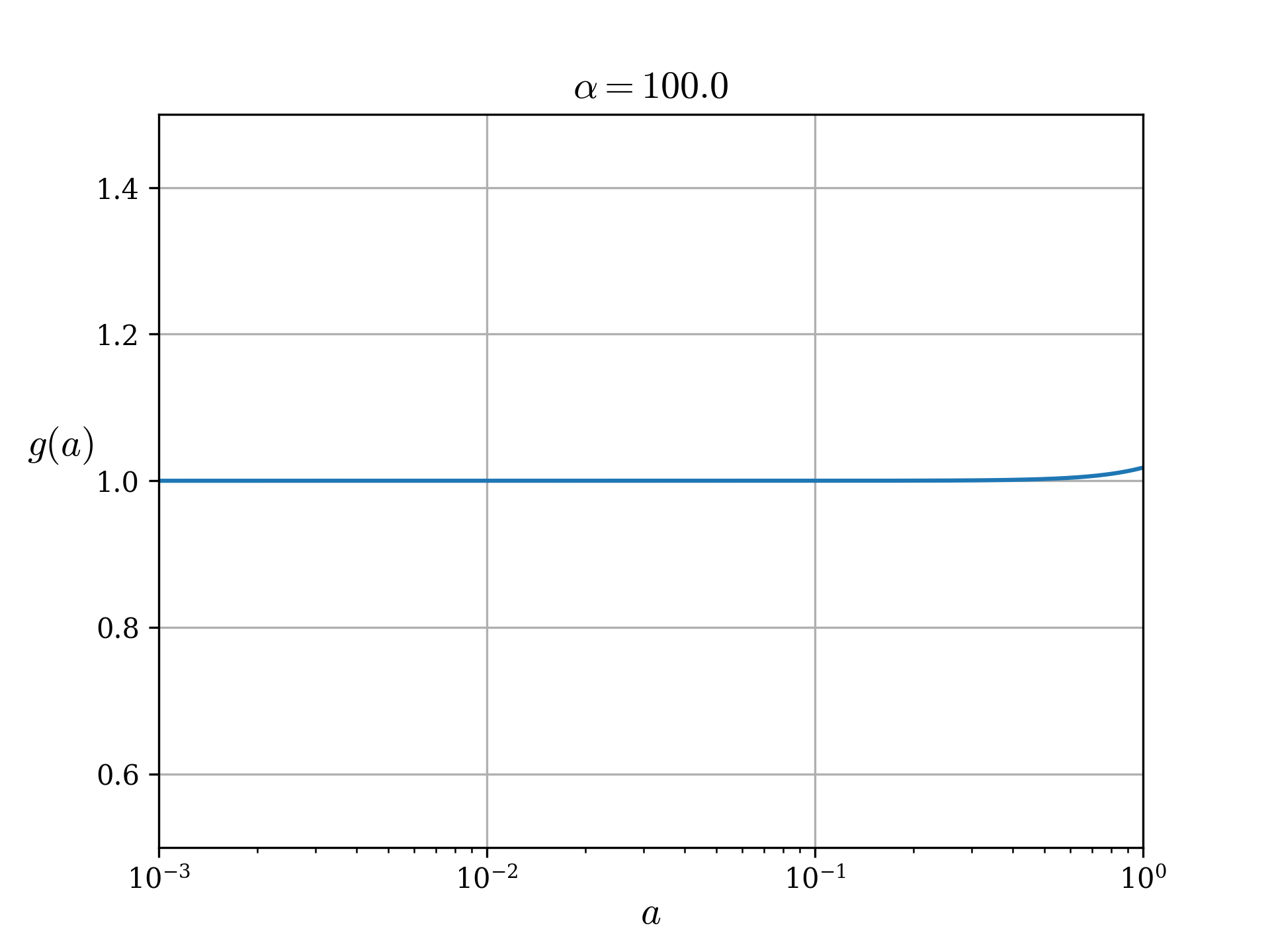}
\includegraphics[scale=0.45]{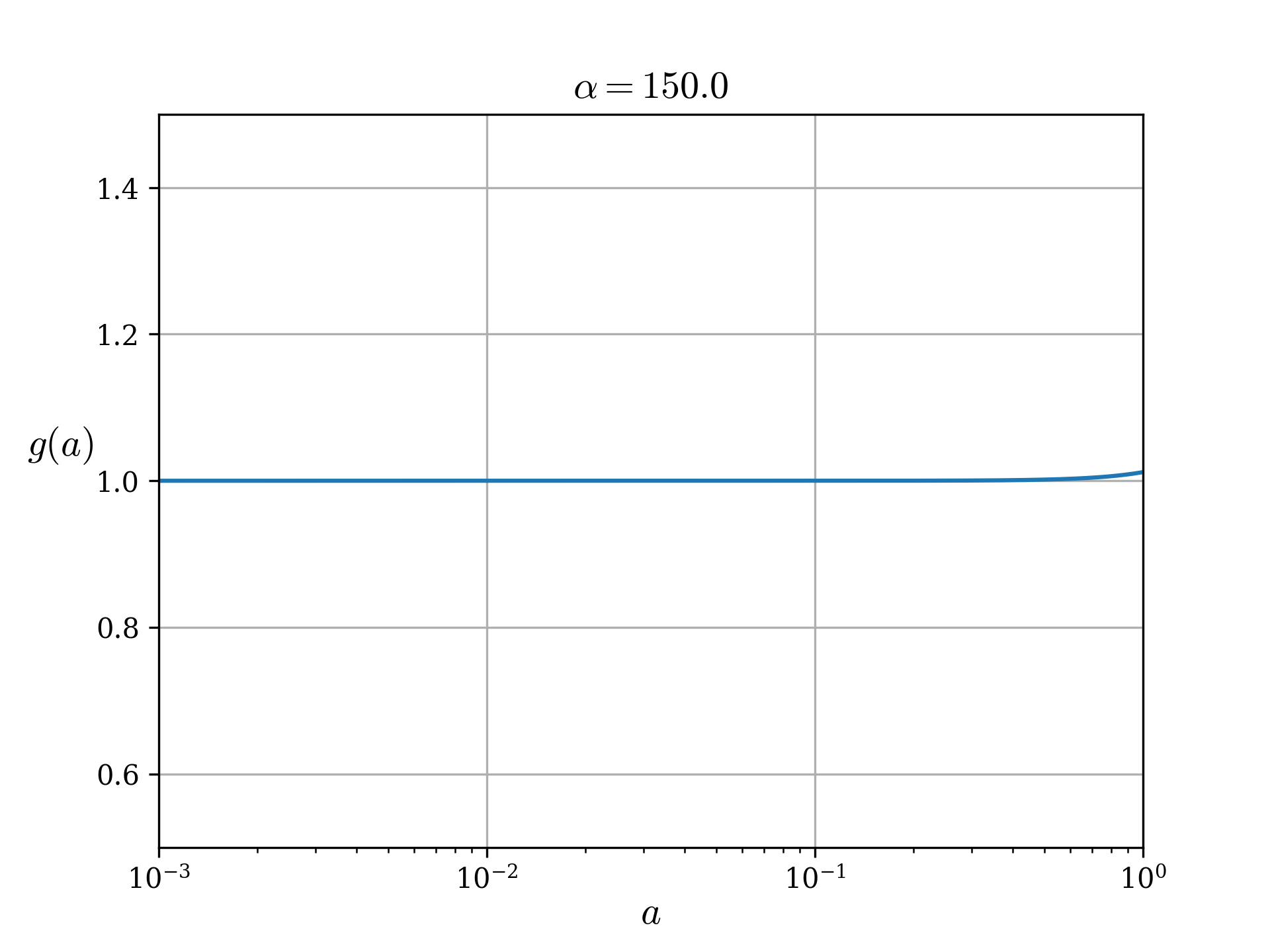}
\includegraphics[scale=0.45]{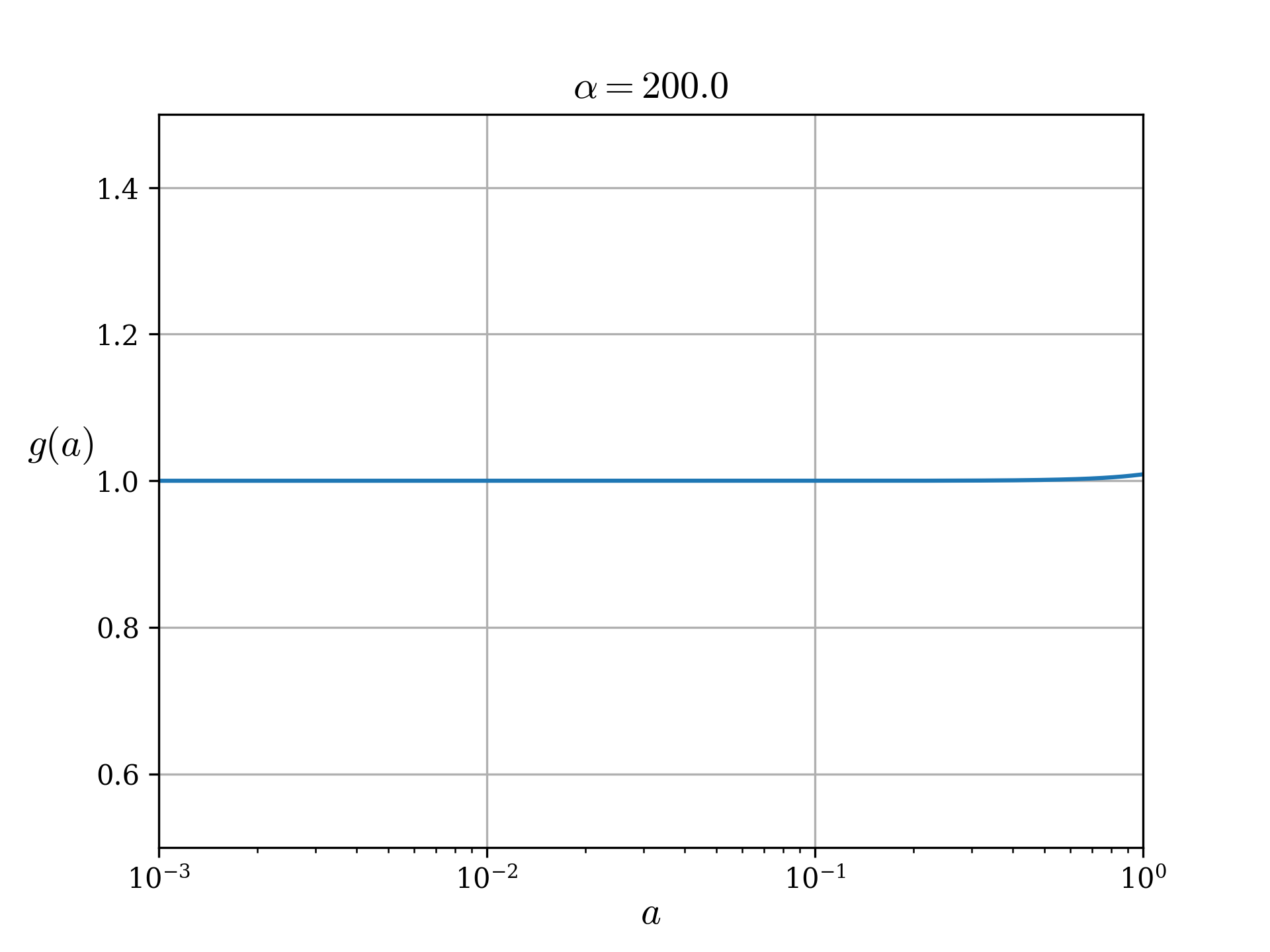}
\caption{Plot of  $g(a)$ as a function of scale factor $a$ for $\alpha = 18, 100, 150$ and $200$ where $g(a)$ is as defined in Eq.~(\ref{Eqn: G(a)}).
It is evident that for large value of $\alpha$, $\rho_{_{DM}} \propto a^{-3}$ where $\rho_{_{DM}}$ is defined in Eq.~(\ref{Eqn:rho_DM})}
\label{Fig: ga}
\end{figure} 
 
With $w_{_{DM}} \approx 0$ and the corresponding $p_{_{DM}} \approx 0$, the energy density of the dark matter part evolves as $\rho_{_{DM}} \propto a^{-3}$.
To confirm this, we introduce the following quantity $g(a)$, defined as
\begin{equation}
g(a) = \l(\frac{\rho_{_{DM}}}{\rho_{_{DMi}}}\r)\l(\frac{a}{a_{_i}}\r)^{3}, \label{Eqn: G(a)}
\end{equation}
where $\rho_{_{DMi}}$ is the dark matter density at an initial epoch $t = t_{_i}$ when $a = a_{_i}$.
The value of $\rho_{_{DMi}}$ can be determined using Eq.~(\ref{Eqn:rho_DM}) together with the initial conditions~(\ref{Eqn: initial conditions}).
In Fig.~(\ref{Fig: ga}), we plot $g(a)$ for four different values of $\alpha$.
When $g(a) = 1$, it implies that $\rho_{_{DM}} \propto a^{-3}$.
For $\alpha = 18$, although $g(a) = 1$ initially, it slightly deviates from unity at the present epoch.
However, for large values of $\alpha$, we find that even at the present epoch, when the potential term has started dominating, $g(a) \approx 1$.
This confirms the fact that the energy density of dark matter part $\rho_{_{DM}}$ defined in Eq.~(\ref{Eqn:rho_DM}) evolves as $\rho_{_{DM}} \propto a^{-3}$ for large values of $\alpha$.

The density parameter $\Omega$ is defined as $\Omega = (8 \pi G \rho)/(3 H^2)$. 
With $\rho_{_{DM}}$ given in Eq.~(\ref{Eqn:rho_DM}) and $\rho_{_{DE}}$ from Eq.~(\ref{Eqn:rho_DE}),  as the energy density of dark matter and dark energy, respectively, the corresponding density parameters  \emph{viz}..\  $\Omega_{_{DM}}$ and $\Omega_{_{DE}}$ turns out to be
\begin{eqnarray}
\Omega_{_{DM}}\,&=&\, \frac{\rho_{_{DM}}}{\rho_{_{DM}} + \rho_{_{DE}} },\label{Eqn: Omega DM}\\
\Omega_{_{DE}}\,&=&\, \frac{\rho_{_{DE}}}{\rho_{_{DM}} + \rho_{_{DE}} }.\label{Eqn: Omega DE}
\end{eqnarray}

\begin{figure}[h] 
\includegraphics[scale=0.40]{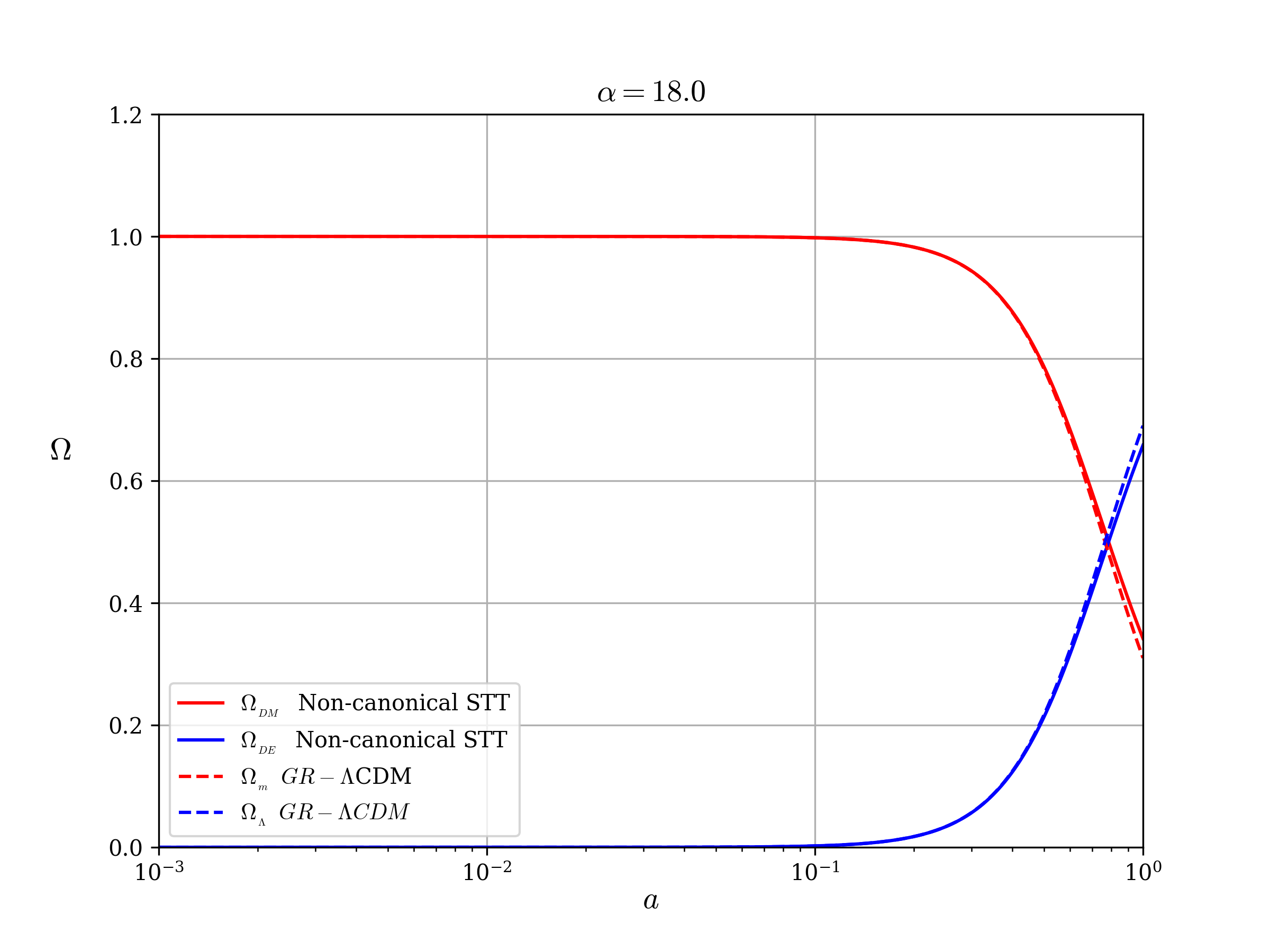}
\includegraphics[scale=0.40]{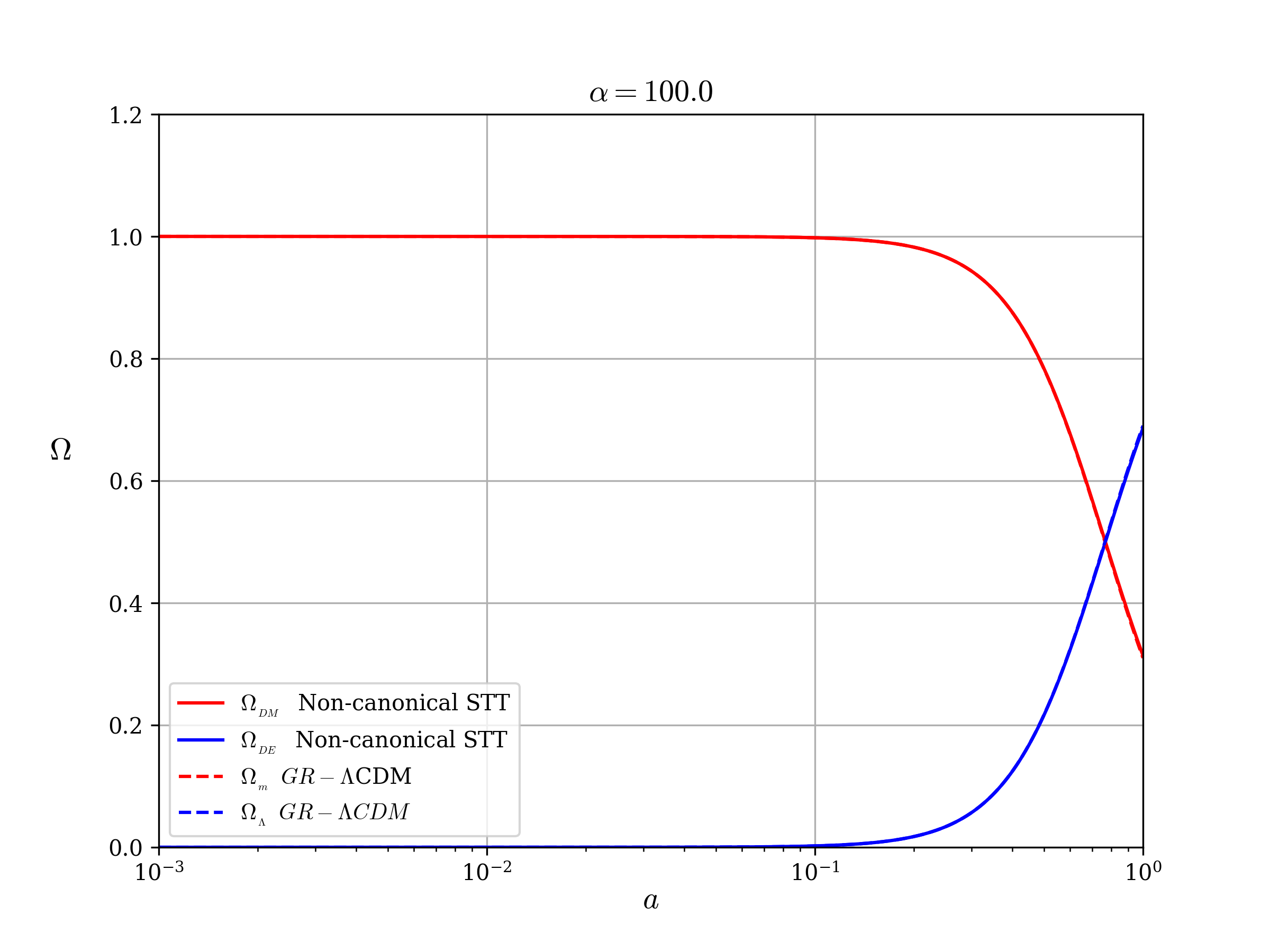}
\caption{The figure shows the plots of density parameters $\Omega_{_{DM}}$ and $\Omega_{_{DE}}$  as a function of scale factor $a$ for $\alpha = 18$ and  $\alpha = 100$.
For comparison, $\Omega_{_m}$ and $\Omega_{_\Lambda}$ of $\Lambda$CDM model is also plotted as dashed lines. 
For large value of $\alpha$, $\Omega_{_{DM}} \approx \Omega_{_{m}}$ and $\Omega_{_{DE}} \approx \Omega_{_{\Lambda}}$ and therefore these curves overlaps when $\alpha = 100$.}
\label{Fig: Omega}
\end{figure}

In Fig.~(\ref{Fig: Omega}), we plot $\Omega_{_{DM}}(a)$ and $\Omega_{_{DE}}(a)$ as a function of scale factor for two different values of $\alpha$.
For comparison, the corresponding    $\Omega_{_{m}}(a)$ and $\Omega_{_{\Lambda}}(a)$ that one gets in the $\Lambda$CDM model based on Einstein's GR is also shown in the same figure.
For $\alpha = 18$, at the present epoch, there is a slight deviation of $\Omega_{_{DM}}(a)$ and $\Omega_{_{DE}}(a)$  from the corresponding  $\Omega_{_{m}}(a)$ and $\Omega_{_{\Lambda}}(a)$, respectively.
However, it is evident from Fig.~(\ref{Fig: Omega}) that for larger values of $\alpha$, $\Omega_{_{m}}(a)$ and $\Omega_{_{\Lambda}}(a)$ evolves just like the GR based $\Lambda$CDM model's $\Omega_{_{m}}(a)$ and $\Omega_{_{\Lambda}}(a)$, respectively.
These results confirms that the model described by the action~(\ref{Eqn: Matterfree action our new NC STT}) mimics the $\Lambda$CDM model.

Finally, it is important to note that in the sclar-tensor theory, the value of the Newtonian gravitational constant $G$ is determined by the scalar field $\phi$.
For example, in the Brans-Dicke theory described by action~(\ref{Eqn: Action BDT}), $G$ is not a constant but evolves as $G \propto \phi^{-1}$.
Similarly, in the non-canonical scalar-tensor theory described by the action~(\ref{Eqn: Action our new NC STT}), $G \propto \phi^{-1}$.
The evolution of $\phi$ in the matter free case described by the action~(\ref{Eqn: Matterfree action our new NC STT}) in a spatially flat Universe are determined by Eq.~(\ref{Eqn: phi dot dot matterfree GNC STT}).
From the solution for $\phi(t)$, it turns out that at the present epoch 
\begin{equation}
\l|\frac{\dot{G}}{G}\r|\,\propto\, \l|\frac{\dot{\phi}}{\phi}\r|\,\approx \, 8.86\times 10^{-12} \mathrm{yr}^{-1}.
\end{equation}
This is well within the bounds from the observations~\cite{Vijaykumar:2020nzc}.

\section{Summary and Conclusions}\label{Sec: conclusions}
In this paper we considered a model of non-canonical scalar-tensor theory described by the action~(\ref{Eqn: Action our new NC STT}).
This can be considered as a non-canonical generalization of the Brans-Dicke theory which corresponds to setting the parameters $\alpha =1$ and $\beta = -1$ in the model.

For an integer value of $\alpha > 1$, we studied the evolution of scale factor $a(t)$ in a spatially flat matter free Universe in this model described by the action~(\ref{Eqn: Matterfree action our new NC STT}) with a linear potential term $V(\phi) = V_{_0}\phi$.

It is shown that when $V_{_0} = 0$ or when the potential term is subdominant, the kinetic term $\lambda X^\alpha \phi^\beta$ leads to a power law solution for the scale factor with $a(t) \propto t^n$.
However, we find the following important results:
\begin{itemize}
  \item[(i)] The maximum value of $n$ in the power law solution $a(t) \propto t^n$ turns out to be $n = (1+\sqrt{3})/4 \approx 0.683$
  \item[(ii)] A solution of the form $a(t) \propto t^{2/3}$ is possible only if the parameter $\alpha \geq 18$.
  \item[(iii)] This solution $a(t) \propto t^{2/3}$ is independent of the choice of the potential $V(\phi)$ provided that potential term is subdominant and the kinetic term drives the evolution of $a(t)$.
\end{itemize}

The solution $a(t) \propto t^{2/3}$ is what one gets in a cold dark matter dominated epoch when the gravity is described by Einstein's GR.
With a choice of linear potential of the form $V(\phi) = V_{_0}\phi$, it is shown that the evolution of $a(t)$ in the model described by the action~(\ref{Eqn: Matterfree action our new NC STT}) 
is the same as one gets in the $\Lambda$CDM model based on Einstein's GR.
This is the main result of this paper.

Although, the recent cosmological observational data indicate a tentative deviation from the $\Lambda$CDM like evolution of the scale factor~\cite{DES:2024jxu, DESI:2025wyn,DESI:2025fii,DESI:2025zgx}, the purpose of this paper is to illustrate that the model with action~(\ref{Eqn: Matterfree action our new NC STT}) can provide a unified description of dark matter and dark energy as if it is a GR based $\Lambda$CDM model.
With an appropriate choice of potential $V\phi)$ other than the linear one, it is possible that the model may provide a unified description of the dark matter and a dynamical dark energy. This needs to be investigated. 
Further, the evolution of the cosmological perturbations, in particular, how the perturbations in the dark matter component evolve, needs to be studied, as this may impose restrictions on the value of the parameters.
\section*{Acknowledgments}
SU acknowledges IUCAA, Pune, India, for providing all the support through the visiting associateship program. SU also thanks the Centre for Theoretical Physics, St. Stephen's College, for the research facilities and support.

\bibliography{ref-ncstt-v2}

\end{document}